\documentclass[hidelinks,10pt,letterpaper]{article}

\usepackage[usenames,dvipsnames,svgnames,table]{xcolor}

\newcommand{\boots}{\mathcal{B}}
\newcommand{\method}{{\tt FPM}}
\newcommand{\reminder}[1]{\textbf{\textsc{\textcolor{red}{(#1)}}}}

\newcommand{\winrv}{\ensuremath{W}}
\newcommand{\outputvar}{\ensuremath{y}}
\newcommand{\inputvec}{\mathbf{x}}

\newcommand{\weight}{\ensuremath{{\bm a}}}

\newcommand{\ratio}{\ensuremath{{r}}}
\newcommand{\success}{\ensuremath{{s}}}
\newcommand{\fieldpos}{\ensuremath{l}}

\usepackage[top=0.85in,left=2.75in,footskip=0.75in]{geometry}
\usepackage{bm}
\usepackage{changepage}

\usepackage[utf8]{inputenc}

\usepackage{textcomp,marvosym}

\usepackage{fixltx2e}

\usepackage{amsmath,amssymb}

\usepackage{cite}

\usepackage{nameref,hyperref}

\usepackage[right]{lineno}

\usepackage{microtype}
\DisableLigatures[f]{encoding = *, family = * }

\usepackage{rotating}
\usepackage{amsmath}
\newtheorem{hyp}{Hypothesis}

\raggedright
\usepackage[aboveskip=1pt,labelfont=bf,labelsep=period,justification=raggedright,singlelinecheck=off]{caption}

\makeatletter
\renewcommand{\@biblabel}[1]{\quad#1.}
\makeatother

\date{}

\usepackage{lastpage,fancyhdr,graphicx}
\usepackage{booktabs}
 \usepackage{multirow}
\fancyheadoffset[L]{2.25in}
\fancyfootoffset[L]{2.25in}
\begin{document}
\vspace*{0.35in}

\begin{flushleft}
{\Large {\tt Footballonomics}: The Anatomy of American Football}\\ Evidence from 7 years of NFL game data
\newline
\\
Konstantinos Pelechrinis\textsuperscript{1}, 
Evangelos Papalexakis\textsuperscript{2}
\\
\bf{1} School of Information Sciences, University of Pittsburgh, Pittsburgh, PA, USA \\
\bf{2} Department of Computer Science, University of California Riverside, Riverside, CA, USA
\\

%
%





* E-mail: Corresponding kpele@pitt.edu
\end{flushleft}

Do NFL teams make rational decisions?
What factors potentially affect the probability of wining a game in NFL?
How can a team come back from a demoralizing interception? 
In this study we begin by examining the hypothesis of {\tt rational coaching}, that is, coaching decisions are always rational with respect to the maximization of the expected points scored.  
We reject this hypothesis by analyzing the decisions made in the past 7 NFL seasons for two particular plays; (i) the Point(s) After Touchdown (PAT) and (ii) the fourth down decisions.   
Having rejected the {\tt rational coaching} hypothesis we move on to examine how the detailed game data collected can potentially inform game-day decisions.  
While NFL teams personnel definitely have an intuition on which factors are crucial for winning a game, in this work we take a data-driven approach and provide quantifiable evidence using a large dataset of NFL games for the 7-year period between 2009 and 2015.  
In particular, we use a logistic regression model to identify the impact and the corresponding statistical significance of factors such as possession time, number of penalty yards, balance between passing and rushing offense etc.   
Our results clearly imply that avoiding turnovers is the best strategy for winning a game but turnovers can be overcome with letting the offense on the field for more time.  
Finally we combine our descriptive model with statistical bootstrap in order to provide a prediction engine for upcoming NFL games.  
Our evaluations indicate that even by only considering a small number of (straightforward) factors, we can achieve a very good prediction accuracy.  
In particular, the average accuracy during seasons 2014 and 2015 is approximately 63\%.  
This performance is comparable to the more complicated state-of-the-art prediction systems, while it outperforms expert analysts 60\% of the time.



\section{Introduction}
\label{sec:intro}

While American football is viewed mainly as a physical game - and it surely is - at the same time is probably the most strategic sports game, a fact that makes it appealing even to international crowd \cite{lamb12}.  
This has lead to people analyzing the game with the use of data analytics methods and game theory.  
For instance, after the controversial last play call of Super Bowl XLIX the Economist argued with data and game theory that this play was rational and not that bad after all \cite{economist-superbowlxlix}.  

The ability to analyze and collect large volumes of data have put forward a quantification-based approach of the success in various sports during the last few years.  
For example, Clark {\em et al.} \cite{clark2013going} analyzed the factors that affect the success of a field goal kick and contrary to popular belief they did not identify any situational factor (e.g., regular vs post season, home vs away etc.) as being significant.  
In another direction the authors in \cite{Pfitzner14} and \cite{warner10} are studying models and systems for determining a successful betting strategy for NFL games, while Stair {\em et al.} \cite{stair08} show that the much-discussed off-field misconduct of NFL players does not affect a team's performance.  
In a different sport, Fewell {\em et al.} \cite{fewell2012basketball} analyzed data from the 2010 NBA play-offs using network theory.  
In particular, they considered a network where each team player is a node and there is an edge between two players if they exchanged a pass.  
Using this structure they found that there is a consistent association between a team's advancement in the next playoff round and the values of clustering and network entropy.  
With regards to soccer, Vilar {\em et al.} \cite{Vilar2013} focus on the analysis of local dynamics and show that local player numerical dominance is key to defensive stability and offensive opportunity.  
Bar-Eli {\em et al.} \cite{bar2007action,bar2009penalty} further collected information from 286 penalty kicks from professional leagues in Europe and South America and analyzed the decisions made by the penalty tackers and the goal keepers.  
Their main conclusion is that from a statistical standpoint, it seems to be more advantageous for a goal keeper to defend by remaining in the goal's center.  
Furthermore, Di {\em et al.} \cite{di2007performance} analyzed the motion of 200 soccer players from 20 games of the Spanish Premier League and 10 games of the Champions League and found that the different positional roles demand for different work intensities.  
In another direction the authors in \cite{Correia2011984} analyzed the pass behavior of rugby players.  
They found that the time period required in order to close the gap between the first attacker and the defender explained 64\% of the variance found in pass duration and this can further yield information about future pass possibilities.  
Predicting the outcome of a sports game has long been a topic of interest and various models have appeared (e.g., \cite{glickman1998state,constantinou2010evaluating,cohea11}).  
However, many of them rely on strong theoretical assumptions (e.g., team strength factors obey to a  first-order autoregressive process) that is hard to verify.  
Data from professional sporting competitions have also been used extensively as a proxy for testing various hypotheses of firm optimization as well as understanding the way people behave and understand - and misinterpret - statistical information (e.g., \cite{risen06,RePEc:ucp:jpolec:v:114:y:2006:i:2:p:340-365,GILOVICH1985295,10.2307/2291254})

Complementary to the existing efforts the purpose of our study is twofold.  
First, we want to examine the {\tt rational coaching} hypothesis.  
In particular, the goal of every team is to win, which is achieved by maximizing the number of points scored.  
Hence, one should expect the decisions that the coaches are making to always be rational with respect to this objective.    
However, triggered by anecdote evidence, such as the article by the Economist mentioned above, it seems that rational decisions made by coaching staff are many times received with questions and are scrutinized.  
Such scrutiny might discourage coaches from following rational decisions especially when they rationale behind its justification is complex.  
The {\em status quo bias} \cite{10.1257/jep.5.1.193} can further perpetuate this behavior and hence, coaches make decisions that differ from those that a rational agent would take to maximize the expected points scored.    
Using game data from the National Football League for the period between 2009 and 2015 we provide convincing evidence to reject the {\tt rational coaching} hypothesis.  
We focus on two particular decisions that the coaches face multiple times during the course of a single game, namely, the Point(s) After Touchdown (PAT) and the fourth down decisions.  
The status quo for PAT is to go for an extra point kick unless if specific situations appear (the supposedly irrelevant factors termed by Thaler \cite{thaler2015misbehaving} that affect human decisions).  
The status quo for fourth down decisions is to punt, unless again if specific situations appear (e.g., game clock running out and trailing in the score).  
Our game data analysis shows that actually (with very few exceptions) both are not the rational decisions to make when the objective is to maximize the expected number of points scored in a game. 
This tenacity of NFL coaches is rather surprising especially given that similar issues have been reported in the literature since 1967 \cite{doi:10.1080/00031305.1967.10479847,sackrowitz2000refining,RePEc:ucp:jpolec:v:114:y:2006:i:2:p:340-365}! 
One might have expected that with data analysis and statistics being an integral part of sports organizations during the last decade, this behavior would have changed and decisions would follow more closely those of an ``ideal'' rational agent.   
However, it appears that the status quo bias is extremely strong in American football.  

Second, 
taking a rational, data-driven approach, we are interested in quantifying the impact of various factors on the probability of wining a game of American football.  
More specifically, 
using the same data as above, we extract specific team statistics for both the winning and losing teams.  
We then use logistic regression to quantify the effect and statistical significance of each of these factors on the probability of wining. 
This model is a descriptive model, i.e., one that analyzes the factors that affect the success of an NFL team.  
However, one of the most intriguing tasks for professional sports analysts is predicting the winners of the upcoming NFL matchups.  
A factor that makes this task particularly hard for American football is the small number of games during a season, which translates to large uncertainty.  
To tackle this problem we propose a combination of statistical bootstrap with the logistic regression model developed above in order to provide a baseline prediction probability.  
While our approach is shown to exhibit an accuracy of approximately 63\% (comparable to that of the state-of-the-art systems such as Microsoft's Cortana), it should be treated as a baseline estimation.    
Simply put the output probability of our model can be though of as an anchor value for the probability of win.  
Further adjustments can be made using information about the specific match-up (i.e., roster, weather forecast etc.).  

In the rest of the study we present the data and methods that we used (see \nameref{sec:materials_methods}).  
We then present the results (see \nameref{sec:results}) and discuss the implications of our study (see \nameref{sec:discussion}). 

\section{Materials and Methods}
\label{sec:materials_methods}

In this section we will present the dataset we used to perform our analysis as well as the different methodological pieces of our analysis.  

{\bf NFL Dataset: }
In order to perform our analysis we utilize a dataset collected from NFL's Game Center for all the games (regular and post season) between the seasons 2009 and 2015.  
We access the data using the Python {\tt nflgame} API \cite{nflgame}.  
The dataset includes detailed play-by-play information for every game that took place during these seasons.  
In total, we collected information for 1,792 regular season games and 77 play-off games.  
Given the small sample for the play-off games and in order to have an equal contribution in our dataset from all the teams we focus our analysis on the regular season games, even though play-off games are by themselves of interest in many perspectives.   

{\bf Mean Field Analysis: }
Many times it is hard - if not intractable at all - to model and examine all the interactions between the components of a complex stochastic system.  
In these cases, we often rely on mean field analysis where the complex interactions are approximated by a single averaged effect.  
Mean field approximations have been extensively used in physics and network science.  
For example, the growth dynamics of various generative network models (e.g., preferential attachment, small-world etc.) have been analyzed through mean field approximations (e.g., \cite{barabasi1999173,newman2000mean}) and provide accurate results despite the abstractions introduced.  
In our work, we will use mean field analysis to estimate the expected point benefit from the various decisions available to a coach.  
This analysis considers interactions between consecutive plays only (e.g., a failed fourth down and the ensuing starting field position for the drive of the opposing team) and ignores higher order interactions (e.g., the potential impact of a decision 5 drives later).

{\bf Bradley-Terry Model for Pairwise Comparisons: }
For modeling the win probability we will use the Bradley-Terry model. 
The Bradley-Terry model is a method for ordering a given set of items based on their characteristics and understanding the impact of these characteristics on the ranking. 
In our case the set of items are the NFL teams and the output of the model for items $i$ and $j$ provides us essentially with the probability of team $i$ (assuming with out loss of generality that it is the home team) winning team $j$. 
In particular, 
the Bradley-Terry model is described by \cite{opac-b1127929}:

\begin{equation}
\Pr(T_i \succ T_j | \pi_i,~\pi_j)=\dfrac{e^{\pi_i-\pi_j}}{1+e^{\pi_i-\pi_j}}
\end{equation}
where $\pi_i$ is the {\em ability} of team $i$. 
Given a set of team-specific explanatory variables $\mathbf{z}_i$, the difference in the ability of the teams $i$ and $j$ can be expressed as:

\begin{equation}
\sum_{r=1}^k \alpha_r (z_{ir}-z_{jr}) + U
\end{equation} 
where $U\sim N(0,\sigma^2)$.  
The Bradley-Terry model is then a generalized linear model that can be used to predict the probability of team $i$ winning team $j$.  
The above formulation does not explicitly treat possible ties between $i$ and $j$ (apart from the fact that if $\Pr(T_i \succ T_j | \pi_i,~\pi_j) = 0.5$ one can consider this as a tie between the two teams).  
However, in our case of NFL game prediction the probability of a game ending with a tie is extremely small and hence, we do not explicitly account for it.  
In particular, in our dataset there are only 3 regular season games that finished with a tie - post season games cannot end with a tie - which corresponds to a 0.1\% probability.  
Nevertheless, there exist extensions of the Bradley-Terry model that are able to deal with ties if this is a highly probable outcome that needs to be modeled \cite{opac-b1127929}.  

{\bf Statistical Bootstrap: }
In order to perform a game outcome prediction, we first need to forecast the performance of each of the contesting teams.  
However, we only have a small set of historic performance data for each team.  
To overcome this problem we will rely on statistical bootstrap.  
Statistical bootstrap \cite{efron93} is a robust method for estimating the unknown distribution of a population's statistic when a sample of the population is known.
The basic idea of the bootstrapping method is that in the absence of any other information about the population, the observed sample contains all the available information about the underlying distribution.
Hence resampling with replacement is the best guide to what can be expected from the population distribution had the latter been available.
By generating a large number of such resamples allows us to get a very accurate estimate of the required distribution.
Furthermore, for data with dependencies (temporal or otherwise), appropriate block resampling retains any dependencies between data points \cite{kunsch89}.
We will utilize bootstrap in our prediction engine.

\section{Results}
\label{sec:results}

In this section we will present our results.  
We will start by examining the {\tt rational coaching} hypothesis and then we will move on to present our logistic regression model.  

\subsection{The {\tt rational coaching} hypothesis}
\label{sec:rational_coaching}

Our running hypothesis is: 

\begin{hyp} 
\label{hyp:h1}{\tt [Rational Coaching]:}
Coaching decisions are always rational in terms of maximizing the number of points scored.  
\end{hyp}

To reiterate, in order to study the above hypothesis we will analyze the decisions taken for the PAT and a fourth down situation.  

\subsubsection{Points After Touchdown}
\label{sec:pat}


Once a team scores a touchdown (worth 6 points) they have the option to either kick an extra point field goal from the 15-yard line (i.e., a 33-yard field goal) or make a regular play from the 2-yard line for 2 points (known as two-point conversion).  
It is clear that this is a strategic decision that teams have to make.  
In some cases the decision is easy (i.e., the team is trailing by 2 points with the clock running out) but in most of the touchdowns the decision is not so clear since special circumstances do not exist. 
In fact, one would expect that such extraordinary situations appear only for the touchdowns towards the end of the game.  
So what about the rest of the touchdowns? 

Analyzing our data we find that the dominant decision is to settle for the extra point after a touchdown instead of trying to score more points and attempting a two-point conversion.  
In particular, from the 9,021 touchdowns during the 7 regular seasons that our data span, only 460 of them were followed by a two-point conversion attempt.  
From these, 235 were successful, that is, an overall 51\% success rate.  
On the other hand, from the 8,561 extra point kick attempts, 8,425 were successful, which translates to a 98.4\% success rate.   
If the decision criteria was the success rate, then the extra point kick is clearly the optimal choice! 
However, a game is won by points and not success rates and therefore, the coaching objective of a rational team should be to maximize the (expected) number of points scored \cite{RePEc:ucp:jpolec:v:114:y:2006:i:2:p:340-365}.  
With $\success_{{\tt 2pts}}$ and $\success_{{\tt kick}}$ being the success rates for the two-point conversions and extra point kicks respectively, the expected point differential benefit per touchdown $\mathbf{E}[p]$ of a two-point conversion over an extra point kick is given by: 

\begin{equation}
\mathbf{E}[p] = 2\cdot \success_{{\tt 2pts}} - 1\cdot \success_{{\tt kick}}
\label{eq:2pts}
\end{equation} 

\begin{figure}[h]
\begin{center}
\includegraphics[scale=0.26]{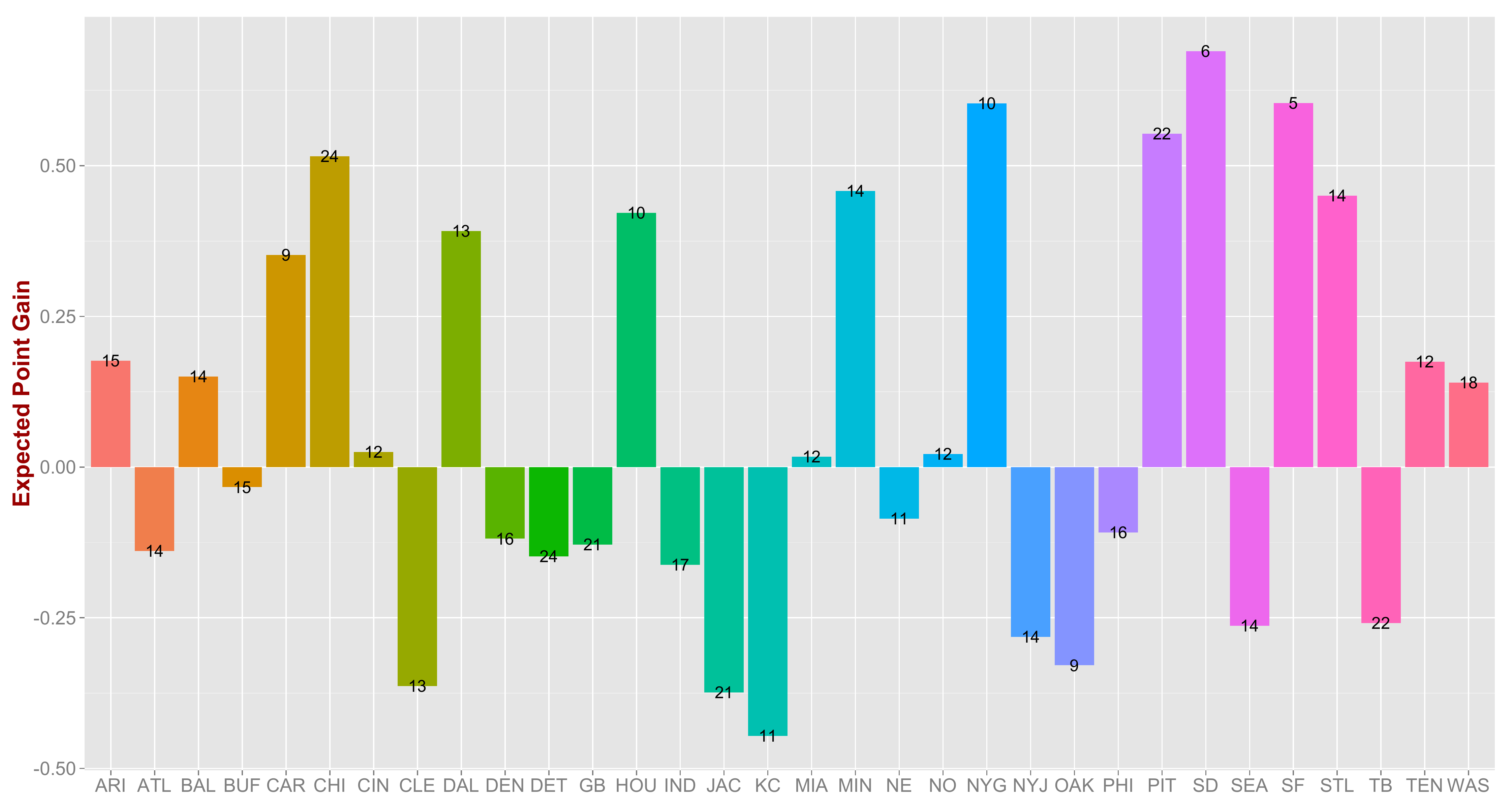}
\caption{Different teams have different expected point gains based on the corresponding success rates $\success_{{\tt2pts}}$ and $\success_{{\tt kick}}$.  On the top of each bar we also present the number of times that the corresponding team attempted a two-point conversion in our dataset.  Pittsburgh Steelers and Chicago Bears appear to have to gain the most from switching to two-point conversions.  In fact, half of the Steeler's two-point conversion attempts in the dataset happened during the 2015 regular season.  The largest expected gain is observed for San Diego.  However, the Chargers have only tried the two-point conversion 6 times and it might be hard to consider the corresponding success rate statistically representative and robust.}
\label{fig:2pts}
\end{center}
\end{figure}

With the extra point kick, the probability of success is 0.984 and thus, the expected number of points is 0.984.  
On the contrary, with the two-point conversion the success rate is only 0.51 but the expected number of points is 1.02, which ultimately gives a positive expected point differential benefit (i.e., $\mathbf{E}[p] > 0$).  
Of course, the net gain per touchdown is fairly small and not all the teams have the same success rate in the two-point conversions and the extra point kick attempts.  
Hence, not all teams would necessarily benefit from this strategy.  
Figure \ref{fig:2pts} presents the expected benefit $\mathbf{E}[p]$ per team using the corresponding success rates from our 7-year dataset.  

Nevertheless, practicing two-point conversion plays can be beneficial in the long term.  
Increasing slightly the success rates can turn to non-negligible point gains.  
In fact, the new rules for the extra point kicks introduced at the beginning of the 2015 season have led to a statistically significant reduction in $\success_{{\tt kick}}$.  
In particular, the extra kick is snapped from the 15-yard line (as alluded to above) instead of the 2-yard line that used to be all the previous years.  
Table \ref{tab:kick_success} depicts the success rates for the extra point kicks over the years that our dataset spans.  
As we can observe there is a significant drop of approximately 5\% ($p$-value $< 10^{-6}$) in the success rate during the 2015 season! 
On the contrary, there was no impact on the success rate of the two-point conversion attempts (as one might have expected).  
The success rate for 2015 was 1.6\% higher than the yearly average success rate between 2009-2014. 
However this difference is not statistically significant ($p$-value = 0.4).   
Therefore, it appears as a good potential strategy for the teams to start practicing and attempting two-point conversions.  

\begin{table}[h]
\scriptsize
  \begin{center}
    \begin{tabular}{l|c|c}
    \toprule
			\bf Year & \bf $\success_{{\tt kick}}$ & \bf $\success_{{\tt 2pt}}$ \\
      \midrule
2009 & 0.9814 & 0.4426\\
2010 & 0.9884 & 0.5471\\
2011 & 0.9942 & 0.5 \\
2012 & 0.9935 & 0.55\\
2013 & 0.9960 & 0.4929\\
2014 & 0.9926  & 0.5161\\
2015 & {\bf 0.9416} & 0.5247\\
\bottomrule
    \end{tabular}
    \vspace{0.2cm}
       \caption{The new PAT rules introduced in 2015 have led to significant drop of the success rate of the extra point kick by approximately 5\% ($p$-value $< 0.001$). On the contrary, there is no impact (as expected) on the two-point conversion success rate.  }
    \label{tab:kick_success}
  \end{center}
\end{table}

Currently, the decision to attempt a two-point conversion depends mainly on the score differential and the time remaining in the game, two supposedly irrelevant factors as per a rational economic model \cite{thaler2015misbehaving}.  
For instance, when a team scores a touchdown which gives them the lead by one point, it is typical to try for the two point conversion that will potentially give a 3 point lead instead of a 2, hence, putting the pressure on the opposing team to score a touchdown to win the game.  
When the touchdown provides a two point lead then things are more complicated and typically teams will attempt the extra point kick.  
What our analysis suggests is that these factors should not impact the decision to attempt a two point conversion, since the expected point benefit is larger regardless of these factors.  
In fact, if a team includes in its playbook the conversion, the success rate $\success_{{\tt 2pts}}$ is also expected to get higher - at least during a transient period - and thus, an even greater benefit is expected.  
Of course at some point the opposing defenses will also adopt and hence, $\success_{{\tt 2pts}}$ will converge to a steady value.    
However, we believe that teams will continue to be conservative, mainly due to the fact that in order for the expected outcome to converge to what is predicted by Equation (\ref{eq:2pts}) a very large number of attempts need to made.  
And while throughout a season this is true, it is not true within a single game where a very small number of attempts will be made by each team (on average 2.5 touchdowns per team per game).  
Nevertheless, our analysis clearly serves as a convincing evidence that coaching decisions are not always rational.  
In fact even teams that have started adopting their game plan - mainly due to the PAT rule changes - are still overly conservative.  
For instance, the Pittsburgh Steelers during the 2015 season attempted for multiple two-point conversion in ``unconventional'' times of the game (e.g., during the first quarter, etc.).  
However, they only attempted 11 conversion (converting 8 of them) out of the 45 touchdowns they scored, which is an overall rate of just about 25\%. 

\subsubsection{Fourth Down Decisions}
\label{sec:4thdown}

Another decision that coaches have to take - more often than the PAT - is what to do in the fourth down situations.  
In American football the teams have 4 tries to advance 10 yards on the field.  
If they fail to do so the opponent takes the ball at the yard line that the team was stopped.  
The teams have to make a choice after the first 3 tries on whether to go for their fourth try and keep their drive alive or whether to punt the ball and push the ensuing drive of the opposing team further from their own goal line.  
Depending on the distance to the goal they might also have the possibility to try for a field goal for 3 points.  

In the vast majority of the cases that coaches face this decision, they decide to either punt or kick a field goal.  
Exceptions of course appear in specific situations, e.g., when a team is trailing by more than 3 points and the clock is running down.  
Using mean field approximations we will calculate the net benefit from ``going for it'' on fourth down.  
In order to estimate the expected benefit we need to compute from our data the following quantities: (i) the conversion rate of a fourth down conversion (as a function of both the field position and the yards to cover for a first down), (ii) the success rate of a field goal (as a function of the distance from the goal) and (iii) the probability of success for a drive (field goal or touchdown) as a function of the starting field position of the drive.  
The latter is needed in order to calculate the (average) impact that a failed fourth down conversion will have on the ensuing drive of the opposing team.  

\begin{figure}[h]
\begin{center}
\includegraphics[scale=0.36]{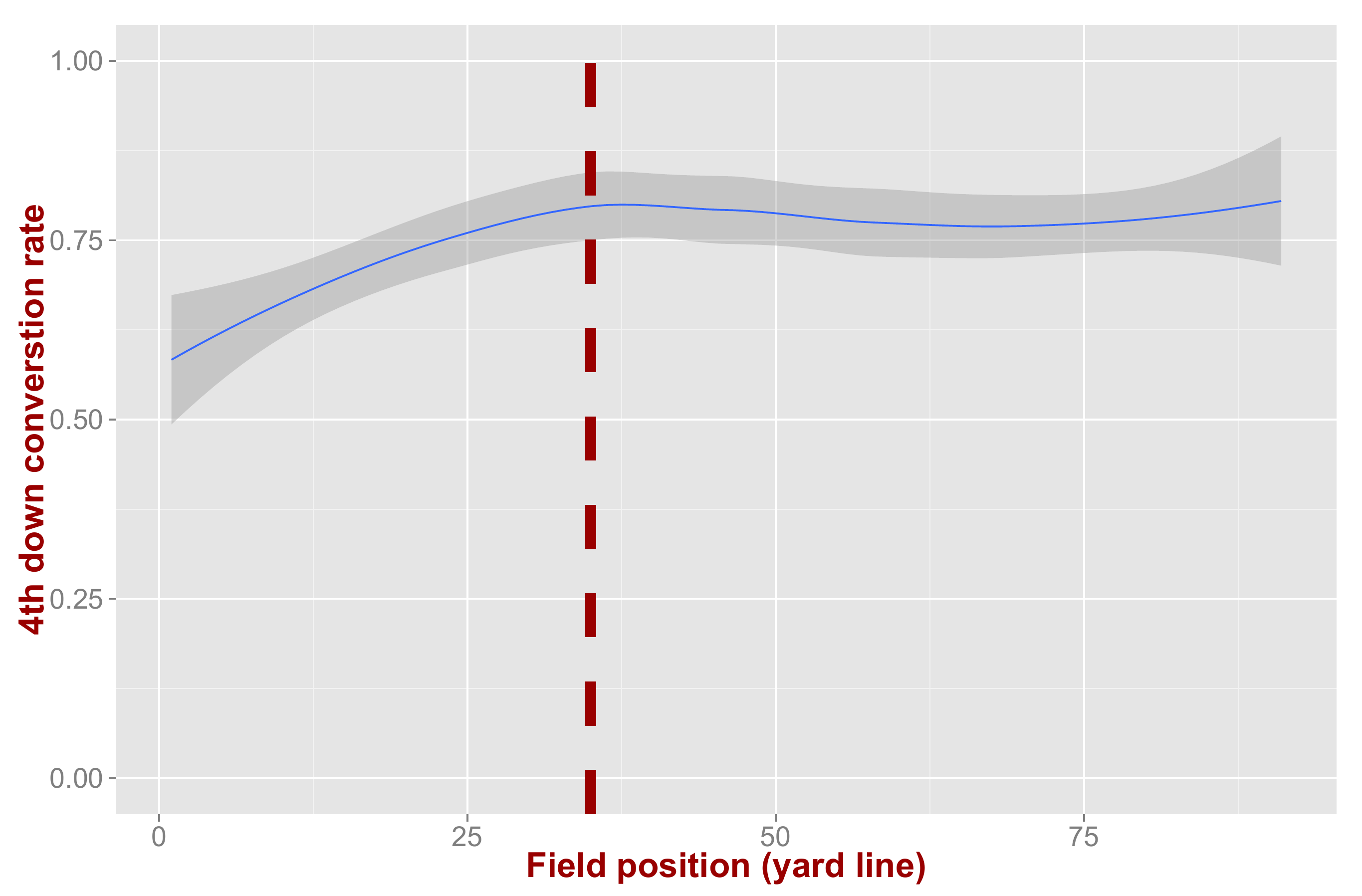}
\caption{The field position of the offense does not seem to impact the outcome of a fourth down conversion, especially beyond the 35-yard line of the offense's territory.}
\label{fig:4th-down-fp}
\end{center}
\end{figure}

{\bf Fourth down conversion rate: }
We begin by examining the success rate of the fourth down conversion attempts.  
Overall, the fourth down conversion rate is a stunning 77.9\%! 
Furthermore, we examine whether this conversion rate is affected by factors such as the position of the offense on the field and the yards remaining for a first down.  
Figure \ref{fig:4th-down-fp} shows the success rate as a function of the field position of the offense. 
As we can see the rate is fairly constant for a long range of field positions.  
In particular, once a team has passed its own 35-yard line it seems that the fourth-down conversion rate is constant.  
It also appears that closer to its own goal line a team has lower conversion rate (always greater than 50\% though).  
However, this might be attributed to the fact that there is an underlying sampling bias in the data in these situations.  
Teams, under normal circumstances (e.g., first quarter drive) do not attempt a fourth down conversion when they are in their own territory, since a failed attempt will give the opponent a very good field position (in fact one that is already in field-goal range).  
The only scenario when a team will attempt a fourth down conversion with this field position (i.e., in their own territory) is a situation where the clock is running down and the team is trailing.  
However, in these cases there is already a lot of time and psychological pressure that the conversion rate is getting a hit.  
In fact, out of the 1,870 fourth down conversion attempts in our dataset, only 202 (i.e., approximately 10\%) occurred while the team was between the 1 and 35-yard lines in its own territory.  
For the same range in the opponent's territory this number was 931 or approximately 49\% of all the attempts.  
Hence, even though there appears to be an impact of the field position on the success rate, we believe that this is more an artifact of the current tactics of the NFL coaching, which leads to sampling bias in our dataset.  

\begin{figure}[h]
\begin{center}
\includegraphics[scale=0.36]{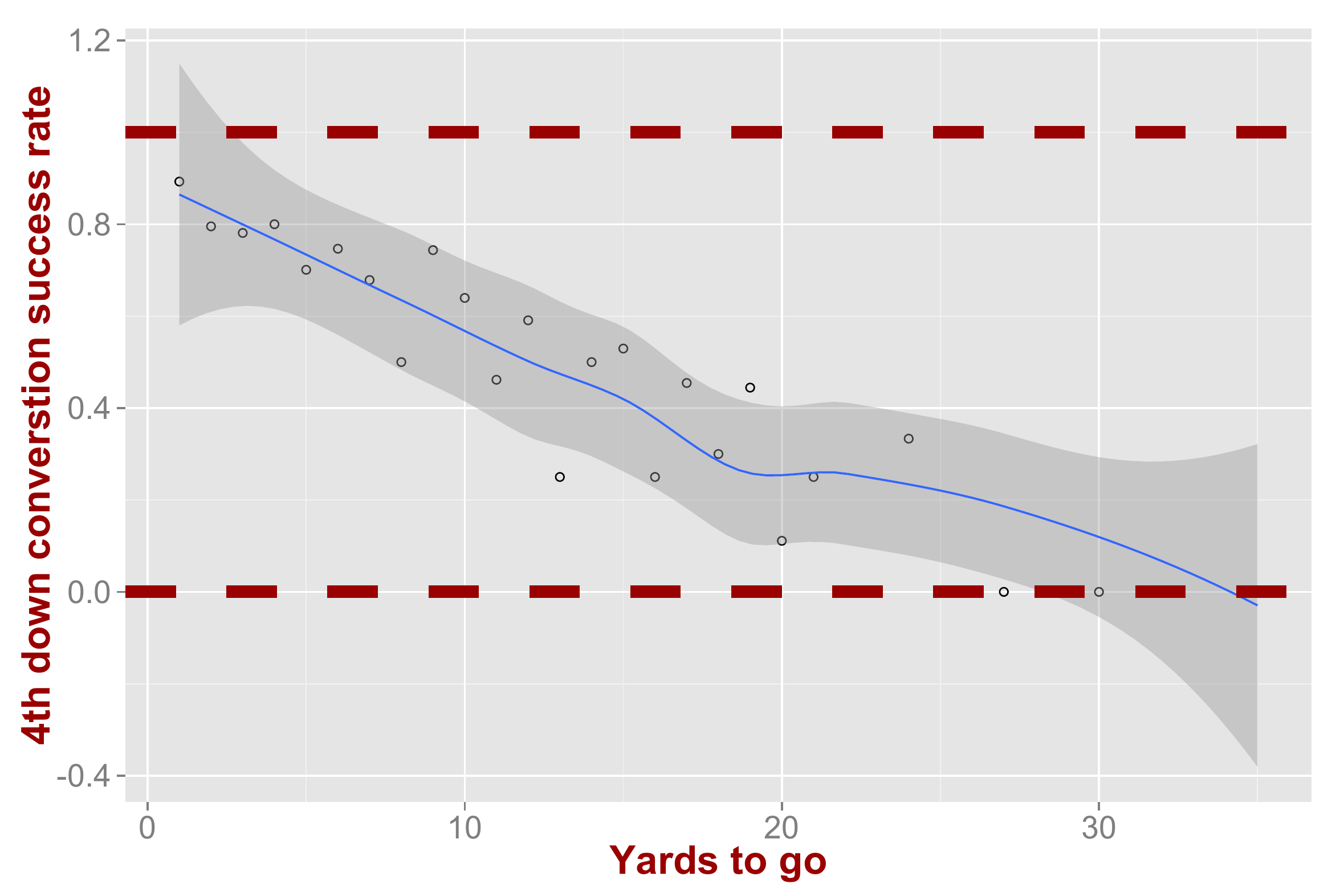}
\caption{The fourth down conversion rate reduces with an increase of the yardage left for the first down.  More than 50\% of the cases that teams go for it on fourth down is for 1 yard.  This skews the overall fourth down conversion rate, since covering more yardage is in general harder as compared to covering 1 yard.  Given that 95\% of the attempts require at most 10 yards to be covered, the adjusted fourth down conversion rate is 73\%.}
\label{fig:4th-down-yardage}
\end{center}
\end{figure}

We further examine the fourth down conversion rate as a function of the yards needed to be covered to get a first down.  
One expects that shorter yardage has a higher success rate as compared to longer yardage.  
Figure \ref{fig:4th-down-yardage} depicts our results.  
In fact, there is a declining trend, that is, the more yards the offense has to cover in the fourth down, the lower the chances of a successful conversion.  
An interesting observation is the fact that 55\% of the fourth down attempts in our dataset are in fourth down and one situations, i.e., the offense has to cover only one yard.  
Hence, the overall fourth down conversion rate is skewed, since the maximum conversion rate is observed in these situations and is equal to 89\%.  
Furthermore, the vast majority of the attempts (95\% of them) require at most 10 yards to reach the first down mark.  
Adjusting for this, we obtain the average of the fourth down conversion rate as 73\%.

\begin{figure}[h]
\begin{center}
\includegraphics[scale=0.26]{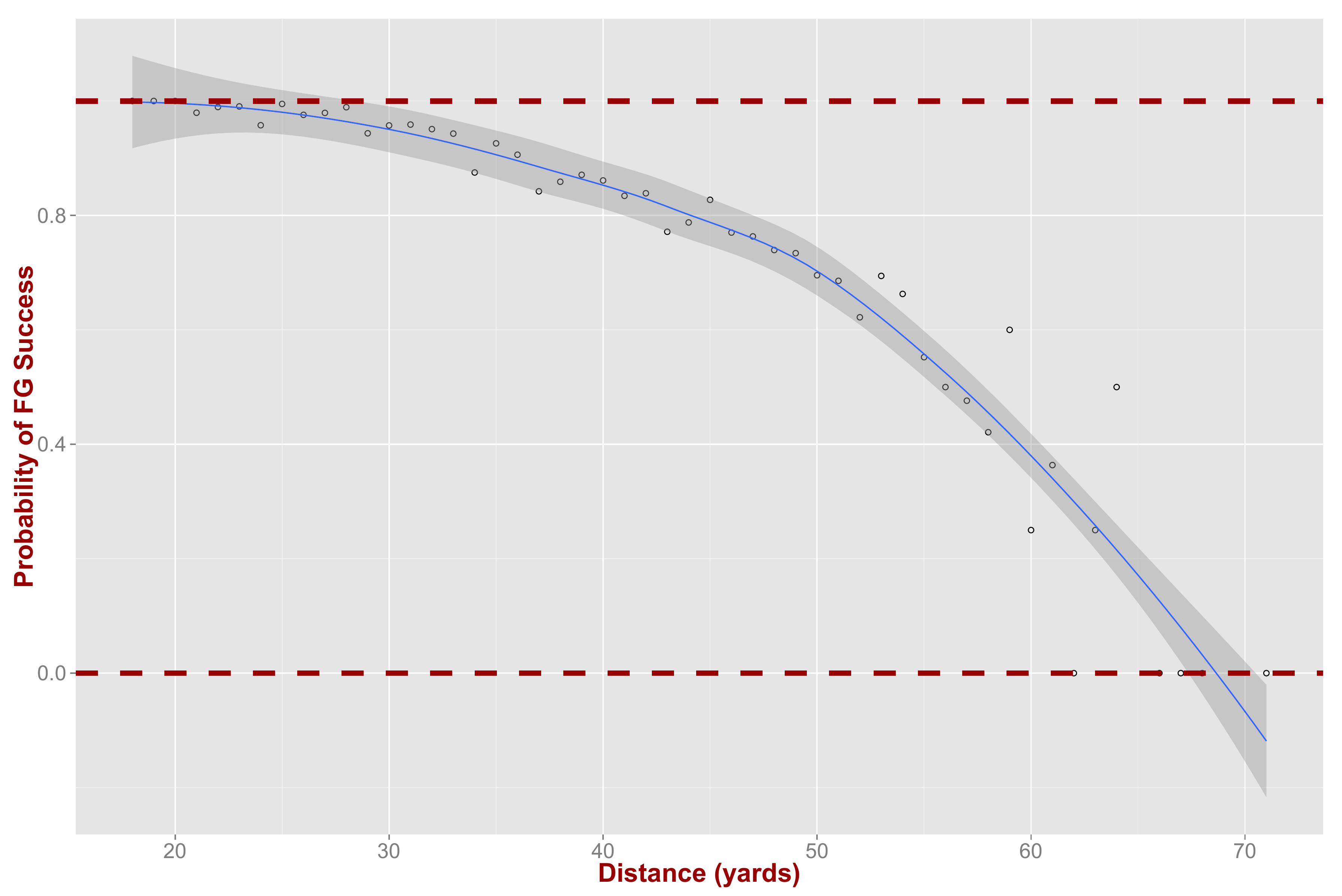}
\caption{The overall success rate of a field goal attempt is 85\%.  The success rate declines with the distance from the goal and exhibits a sharp decline beyond a distance of 50 yards.  This sharp decline might be simply an artifact of the fewer data points in this regime (only 11\% of the field goal attempts were beyond a 50 yard distance).}
\label{fig:fg-dist}
\end{center}
\end{figure}

{\bf Field Goal Success Rate: }
When teams face a fourth down decision in their opponent's territory they can decide to settle for a field goal, which will provide them with 3 points if successful.    
In order to calculate the expected payoff from a field goal attempt we calculate the success rate of the kick as a function of the distance from the goal.  
Figure \ref{fig:fg-dist} depicts our results.  
As we can see there is a slowly declining field goal success rate, which exhibits a steeper decline after the 50 yards.  
However, only 11\% of the field goal attempts in our dataset had a distance larger than 50 yards. 
These typically correspond to efforts to tie or win a game when the clock is running down; under regular circumstances teams would most probably have punted the ball.  
Overall, the success rate of a field goal (not controlling for the distance from the goal) is 85.5\%.

{\bf Success of a Drive and Starting Field Position: }
Turning the ball on downs (i.e., failing a fourth down conversion) does not only impact the current drive of the offense by terminating it with 0 points, but it might also give the opponent a very good starting field position.  
After a touchdown or a field goal the team that scored, kicks the ball from their own 35 yard line.  
The opponent has a dedicated player (called kick returner) who receives the ball and tries to advance it on the field.  
Typically, most of the kicking teams attempt to kick the ball beyond the opponent's goal line, which will not give the opportunity to the kick returner to advance the ball.  
In this case the next drive starts from the offense's 20 yard line (i.e., the offense has to cover 80 yards to score a touchdown).  

\begin{figure}[h]
\begin{center}
\includegraphics[scale=0.36]{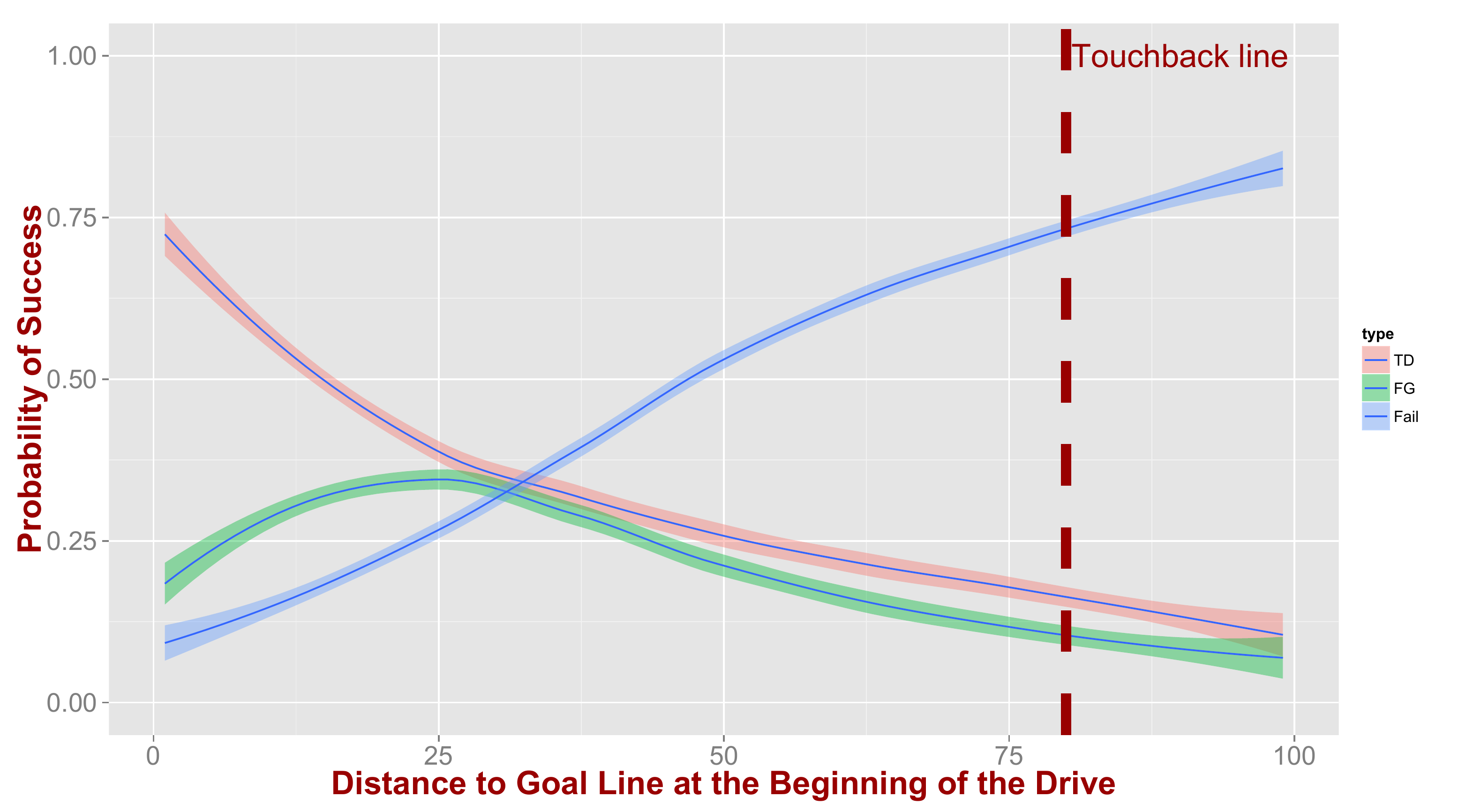}
\caption{A drive that starts in the opponents territory has significantly increased probability of leading to a scoring play (touchdown or field goal).  On the contrary, turning the ball on downs at the opponents territory (between the yard lines 50 and 20) increases the probability of the ensuing drive to lead to a score by only 7\% on average as compared to the baseline case of a drive starting after a touchback. }
\label{fig:drive-fp}
\end{center}
\end{figure}

The starting position of a team can potentially impact the success of the drive.  
If the kick returner is able to advance the ball well beyond his own 20 yard line (e.g., at the opponent's territory) then the offense has a short field to work with and increased chances of scoring.  
The same is true if the team gets the ball well beyond their own 20 yard line due to a turn on downs from the opponent.  
Hence, in order to calculate the potential loss from a failed fourth down conversion we need to estimate how a turn on downs will impact the success of the ensuing drive from the opposing team.  
Figure \ref{fig:drive-fp} presents the fraction of drives that resulted in a field goal, touchdown or failed (i.e., ended with a turnover or a punt) as a function of the starting position captured by the distance that needs to be covered to score a touchdown.  
As we can see when the distance that the offense has to cover at the beginning of the drive is less that 25 yards the probability of scoring a touchdown is rapidly increasing, while the probability of not scoring at all reduces rapidly as well. 
However, both of them are much smaller compared to the probability of a failed drive (for a distance greater than 50 yards).

Clearly the starting field position can impact the success of the drive especially when this drive starts at the opponents territory.  
This might be the reason that an offense rarely (under regular game situations) goes for fourth down when they are in their own territory; a failed conversion increases dramatically the chances for a scoring ensuing drive from the opponent.  
However, when the offense has entered the opponents territory (i.e., beyond the 50 yards in the figure) failing to convert on a fourth down increases the chances of a scoring ensuing drive by only 7\% on average as compared to the baseline case of a touchback.

We will make use of these results and curves in what follows to estimate the mean field approximation of the net gain for going for it on a fourth down situation.  
A negative net gain will provide supporting evidence for the {\tt rational coaching} hypothesis, while a positive one will provide further evidence against it.  

\vspace{0.2in}
{\bf Mean Field Net Point Benefit}

For the mean field approximation of the net point gain $\mathbf{E}[P]$ we need to calculate the expected point benefit $\mathbf{E}[P^+]$ from a successful fourth down conversion as well as the expected point cost $\mathbf{E}[P^-]$ from a potential failed conversion.  
The expected benefit $\mathbf{E}[P^+]$ is a function of the fourth down conversion rate $\success_{{\tt 4conv}}$ as well as the field position $\fieldpos$, i.e.,  $\mathbf{E}[P^+] = f_1(\success_{{\tt 4conv}}, \fieldpos)$.  
On the contrary, the expected point cost $\mathbf{E}[P^-]$ is a function of the success rate of a field goal $\success_{{\tt fg}}$, which itself is a function of the field position $\fieldpos$, and the increase in the probability $\Delta \pi_{{\tt td}}$ and $\Delta \pi_{{\tt fg}}$ of the ensuing opponent's drive leading to a touchdown or field goal score respectively, which itself depends on the field position $\fieldpos$ as well, i.e., $\mathbf{E}[P^-] = f_2(\success_{{\tt fg}},\Delta \pi_{\success},\fieldpos)$.  
In particular, 

\begin{equation}
\mathbf{E}[P^+]= 6 \cdot \success_{{\tt 4conv}}^{\gamma(\fieldpos)} 
\label{eq:mfa1}
\end{equation}

\begin{equation}
\mathbf{E}[P^-]= 3 \cdot \success_{{\tt fg}} + (3\cdot  \Delta \pi_{{\tt fg}}+ 6\cdot  \Delta \pi_{{\tt td}})
\label{eq:mfa2}
\end{equation}
where $\gamma(\fieldpos)$ is the number of times that the offense will need to convert a fourth down to reach the goal line.  
Clearly the further from the goal line (i.e., small $\fieldpos$) the larger the expected value of $\gamma(\fieldpos)$.  
In order to have a realistic estimate for $\gamma(\fieldpos)$ we analyzed all the approximately 43,000 drives from all the games in our dataset.  
The average drive length is 29 yards.  
This means that, on average in order to keep the drive alive, the team will need to convert on fourth down once every 29 yards.  
Therefore, if a team starts at its own 20-yard, this means that they will have to successfully convert on average 2.7 times before reaching the goal line.  
In general, with $\fieldpos$ being the starting field position of a team (i.e., the yards to cover are $100-\fieldpos$) we have $\gamma(\fieldpos) = \dfrac{100-\fieldpos}{29}$.  

\begin{figure}[h]
\begin{center}
\includegraphics[scale=0.36]{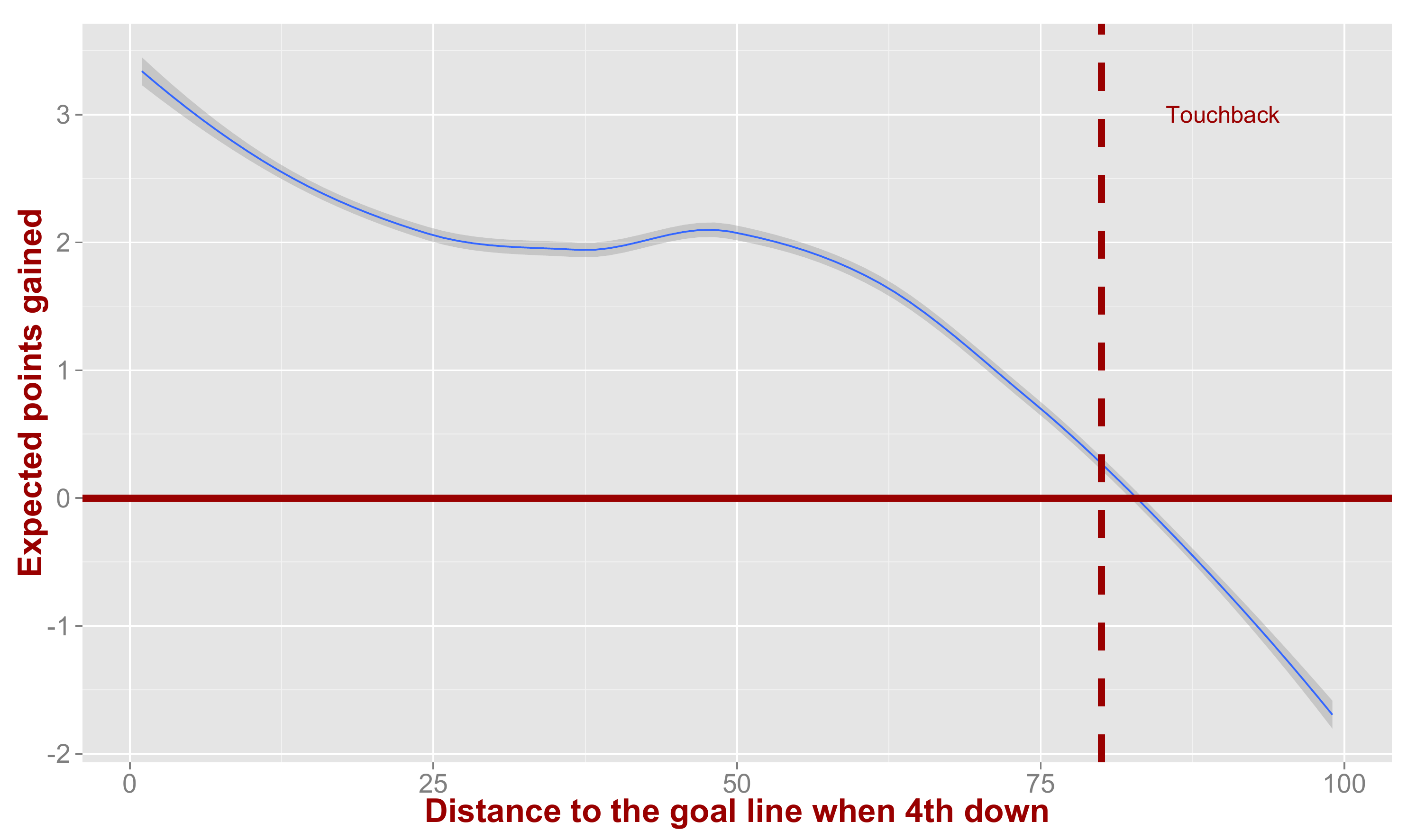}
\caption{The expected point benefit is positive for most of the field length.  The benefits become particularly tempting, i.e., $\mathbf{E}[P] > 2$, when the offense has stepped into the opponent's territory.  Overall, the expected point benefit is 1.4 ($p$-value $<$ 0.01) and the fact that teams typically punt or settle for a field goal on fourth down provides additional evidence for rejecting the {\tt rational coaching} hypothesis.  }
\label{fig:point_ben}
\end{center}
\end{figure}

\begin{figure}[h]
\begin{center}
\includegraphics[scale=0.36]{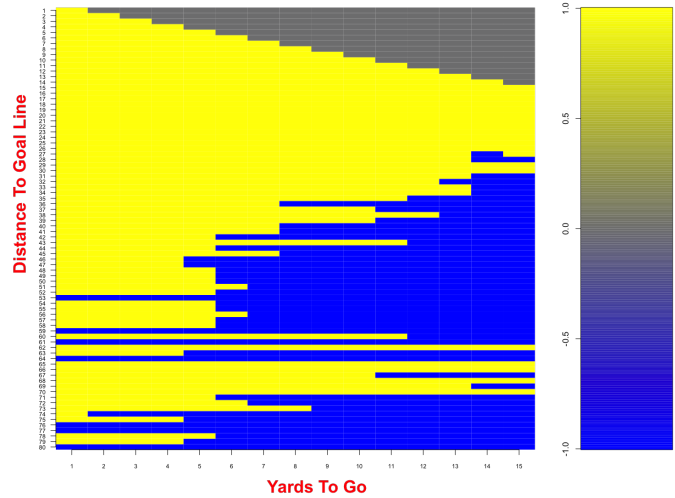}
\caption{A cheat sheet for the fourth down decision. }
\label{fig:cheat-sheet}
\end{center}
\end{figure}

Using the results from Figures \ref{fig:4th-down-fp}, \ref{fig:fg-dist} and \ref{fig:drive-fp} we obtain Figure \ref{fig:point_ben}. 
This figure depicts the expected point benefit (i.e., $\mathbf{E}[P] = \mathbf{E}[P^+] - \mathbf{E}[P^-]$) that the offense will have as a function of the field position at the first time it faces a fourth down situation.  
In particular, the horizontal axis represents the distance to the goal line.  
As we can see, $\mathbf{E}[P]$ is positive for more than 80\% of the field.  
In fact the average point gain is 1.4 points per drive ($p$-value $<$ 0.01), which can translate to significant point gains over the course of a game.  
Finally, Figure \ref{fig:cheat-sheet} depicts a decision chart for making the decision to go for it on a fourth down situation based on our mean field approximation.  

In conclusion, even though there clearly are cases where going for it on fourth down does not provide any benefit - i.e., in the regime where $\mathbf{E}[P] < 0$ - teams are extremely reluctant on attempting the conversion even though the game data show that there can be significant point benefits.  
Our analysis with respect to the fourth down decisions (and in particular the results in Figure \ref{fig:point_ben} from our mean field approximation) further supports the rejection of the {\tt rational coaching} hypothesis.  

\subsection{Game Day Models}
\label{sec:model}

In this part of our study we will present our descriptive generalized linear model.  
In particular, we build a Bradley-Terry model to understand the factors that impact the probability of a team winning an American football game.  
This model will be later used in our future matchup prediction engine, {\method}, as we describe in Section \ref{sec:prediction}.

Let us denote with $\winrv_{ij}$ the binary random variable that represents the event of home team $i$ winning the game against visiting team $j$.  
$\winrv_{ij} = 1$ if the home team wins the game and 0 otherwise.  
As aforementioned our model for $\winrv_{ij}$ will provide us with the probability of the home team winning the game given the set of input features, i.e., $\outputvar=\Pr(\winrv_{ij} = 1 | \inputvec)$.  
The input of this model is vector $\inputvec$ that includes features that can potentially impact the probability of a team winning.  
The features we use as the input for our model include: 

{\bf  Total offensive yards differential: }
This feature captures the difference between the home and visiting teams' total yards (rushing and passing) produced by their offense in the game.  

{\bf Penalty yards differential: }
This features captures the differential between the home and visiting teams' total penalty yards in the game.  

{\bf Turnovers differential: }
This feature captures the differential between the total turnovers produced by the teams (i.e., how many times the quarterback was intercepted, fumbles recovered by the opposing team and turns on downs).  


{\bf Possession time differential: }
This feature captures the differential between the ball possession time between the home and visiting team. 

{\bf Passing-to-Rushing ratio $\ratio$ differential: }
The passing-to-rushing ratio $\ratio$ for a team corresponds to the fraction of offensive yards gained by passing: 

\begin{equation}
\ratio = \dfrac{\textsf{\# of passing yards}}{\textsf{\# of total yards}}
\label{eq:ratio}
\end{equation}

This ratio captures the offense's balance between rushing and passing.  
A perfectly balanced offense will have $\ratio=0.5$. 
We would like to emphasize here that $\ratio$ refers to the actual yardage produced and not to the passing/rushing attempts.  
The feature included in the model represents the differential between $\ratio_{\tt home}$ and $\ratio_{\tt visiting}$.


{\bf Power ranking differential: }
This is the current difference in rankings between the home and the visiting teams.  
A positive differential means that the home team is {\em stronger}, i.e., ranks higher, than its opponent.  
For the power ranking we utilize {\tt SportsNetRank} \cite{sportsnetrank}, which uses a directed network that represents win-lose relationships between teams.  
{\tt SportsNetRank} captures indirectly the schedule strength of a team and it has been shown to provide a better ranking for teams as compared to the simple win-loss percentage.

\begin{figure}[!h]
\centering
\includegraphics[scale=0.15]{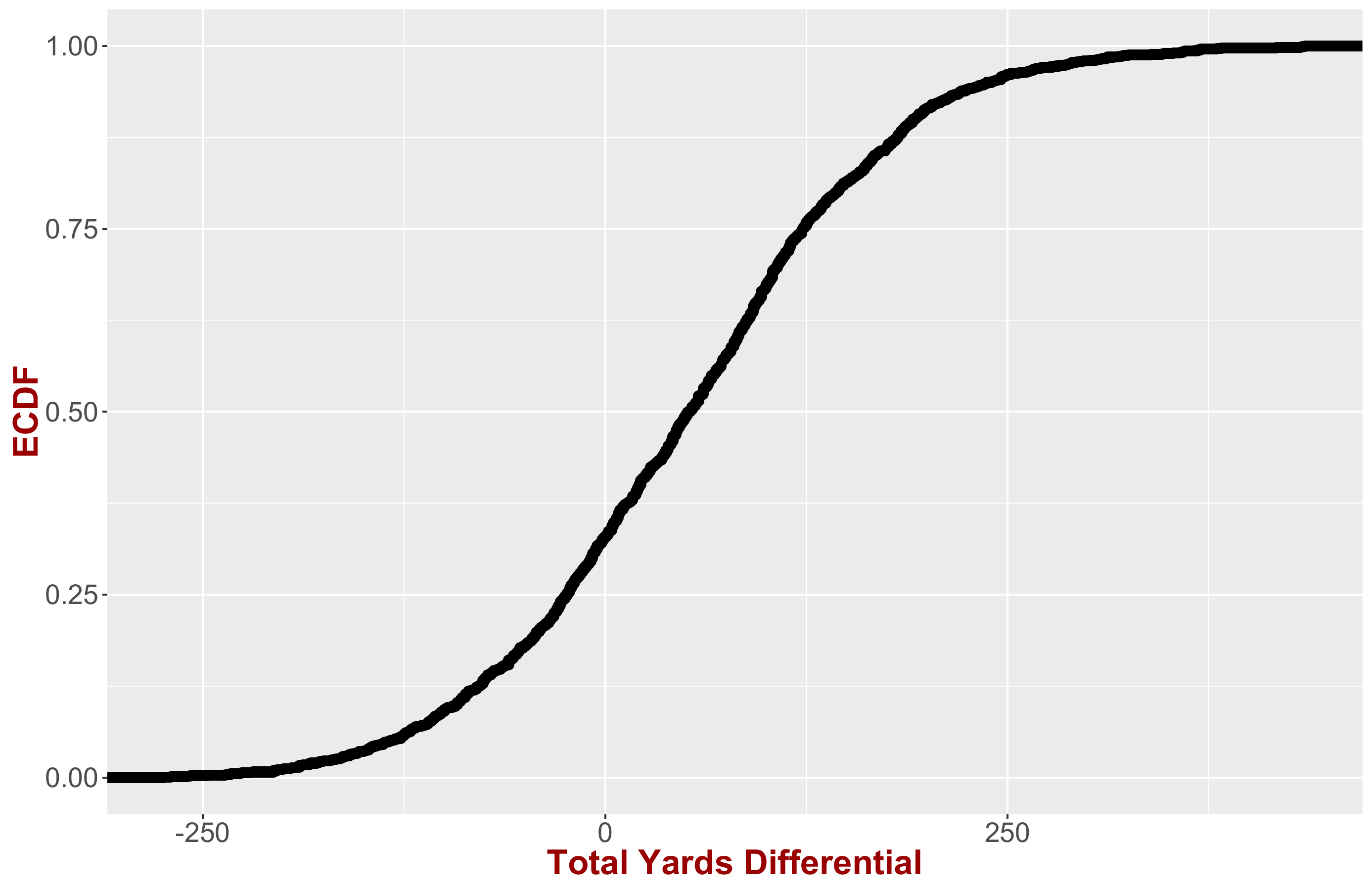}\includegraphics[scale=0.15]{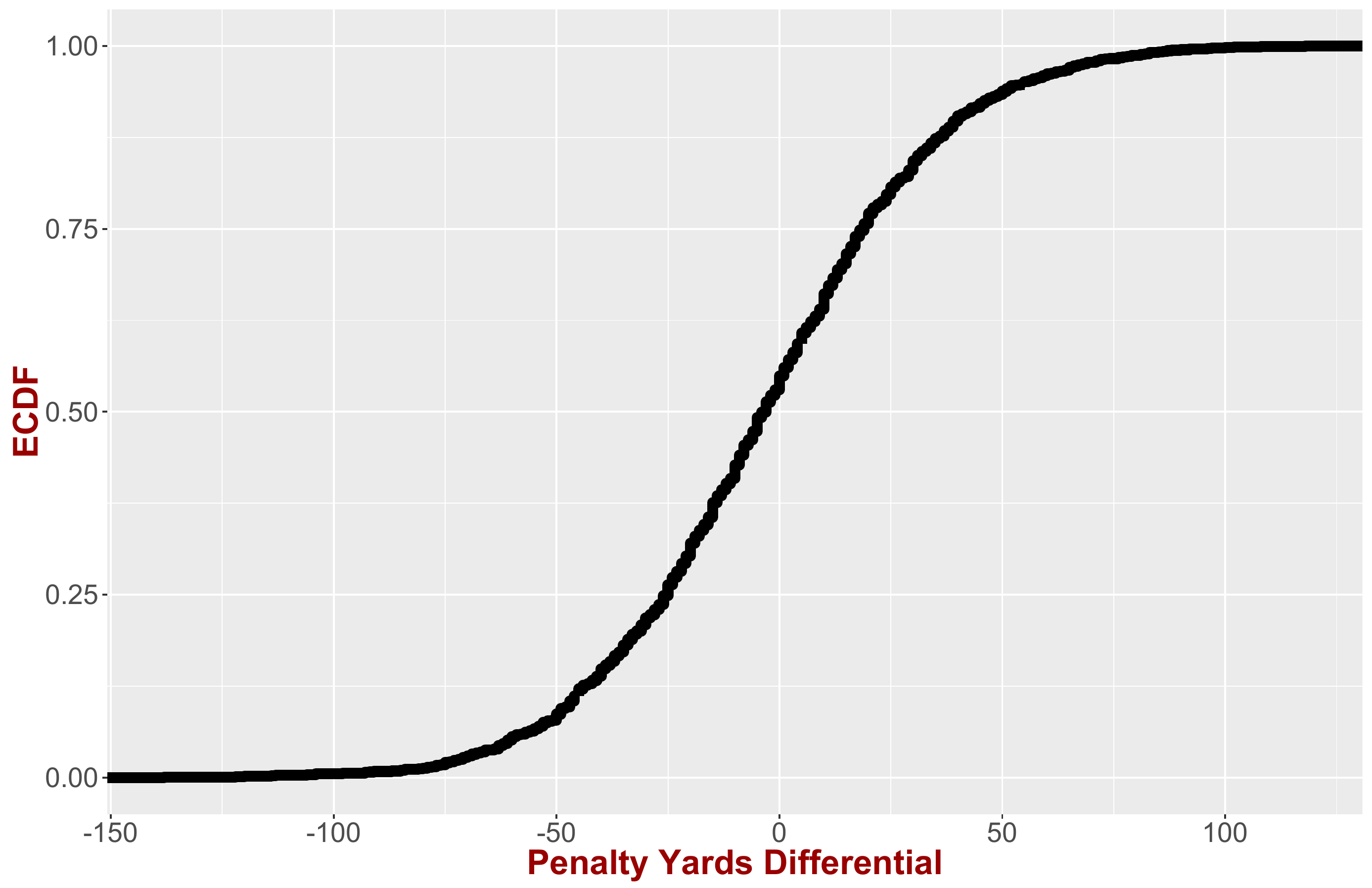}\includegraphics[scale=0.15]{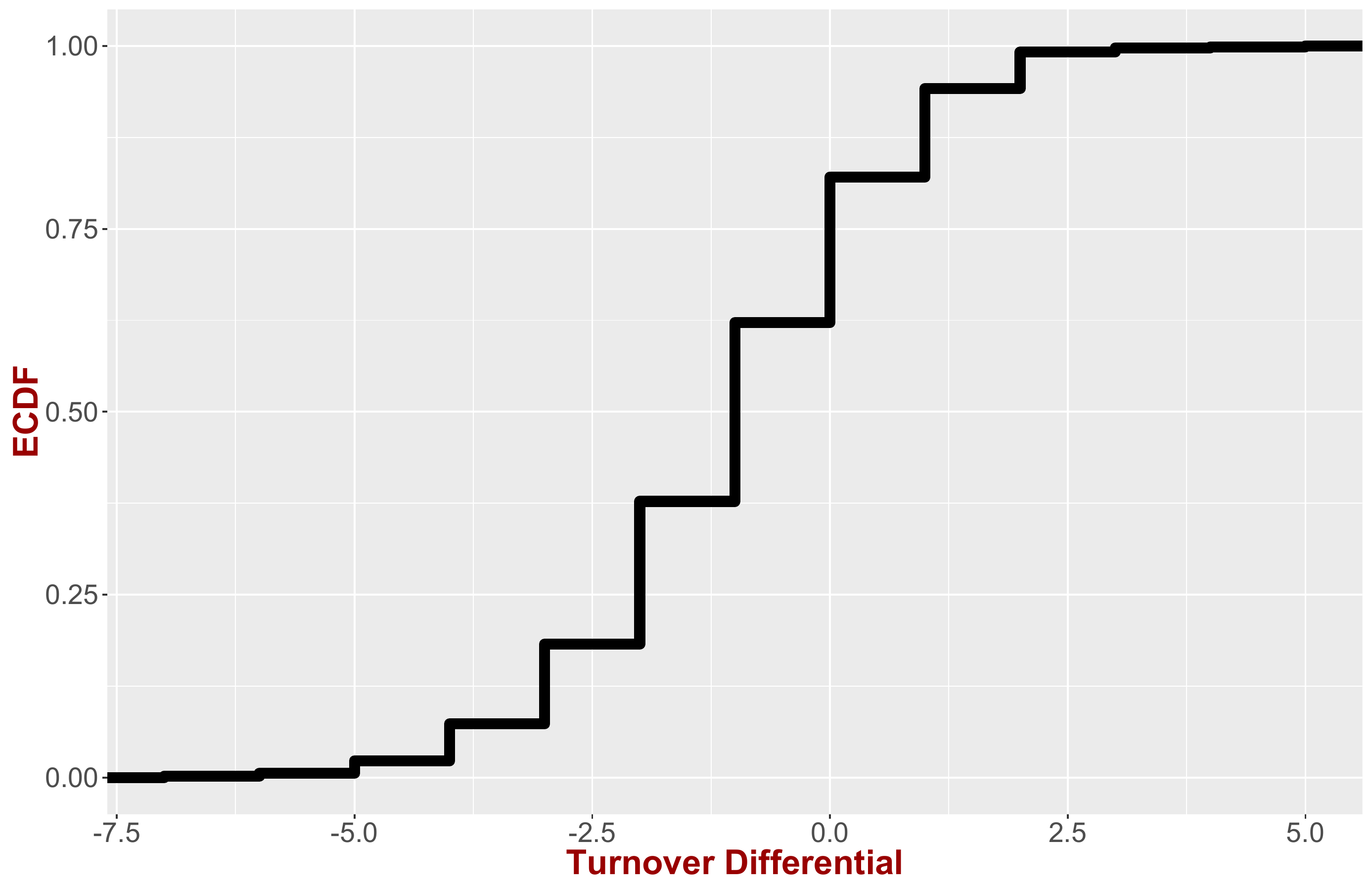}
\includegraphics[scale=0.15]{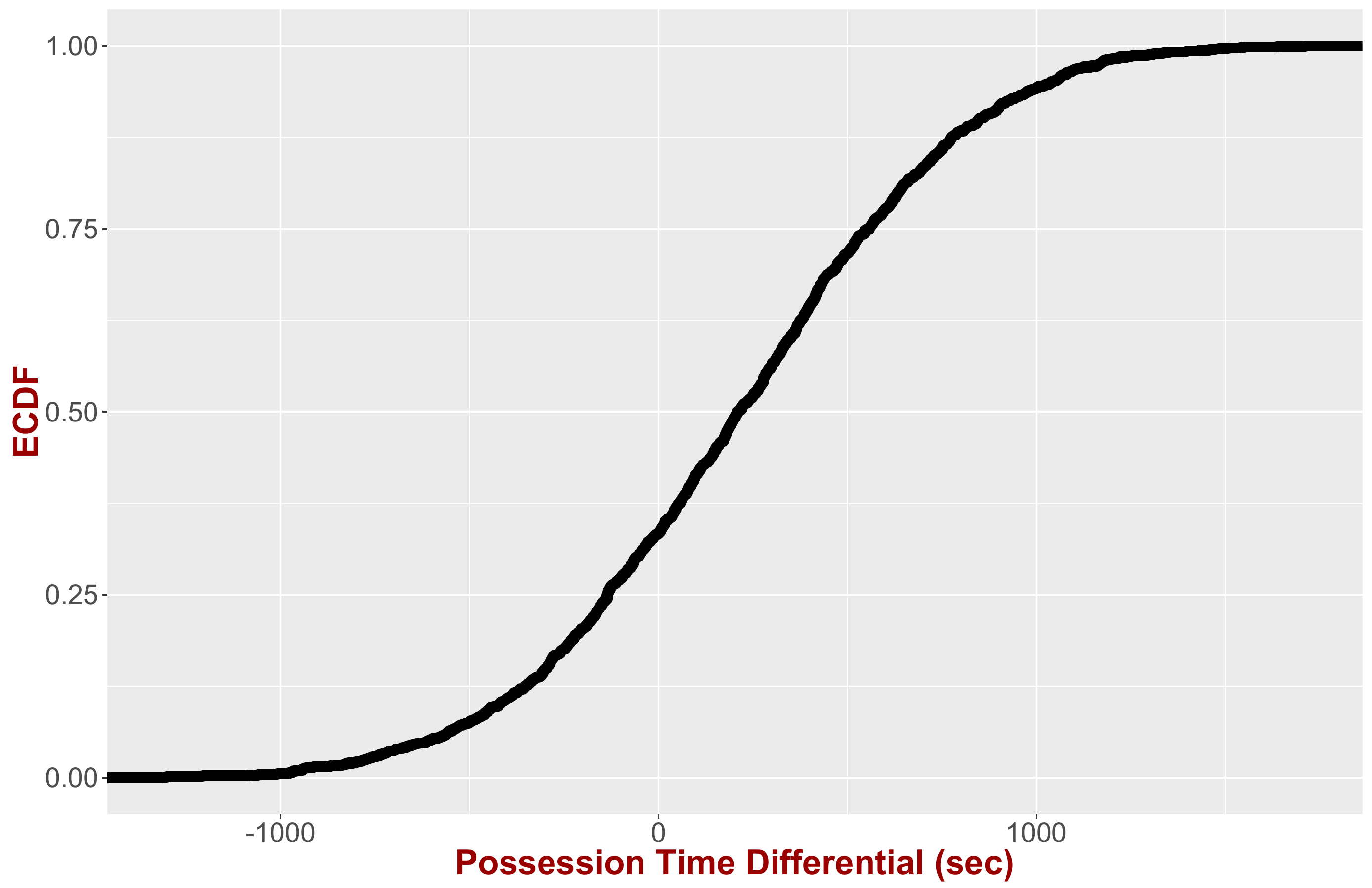}\includegraphics[scale=0.15]{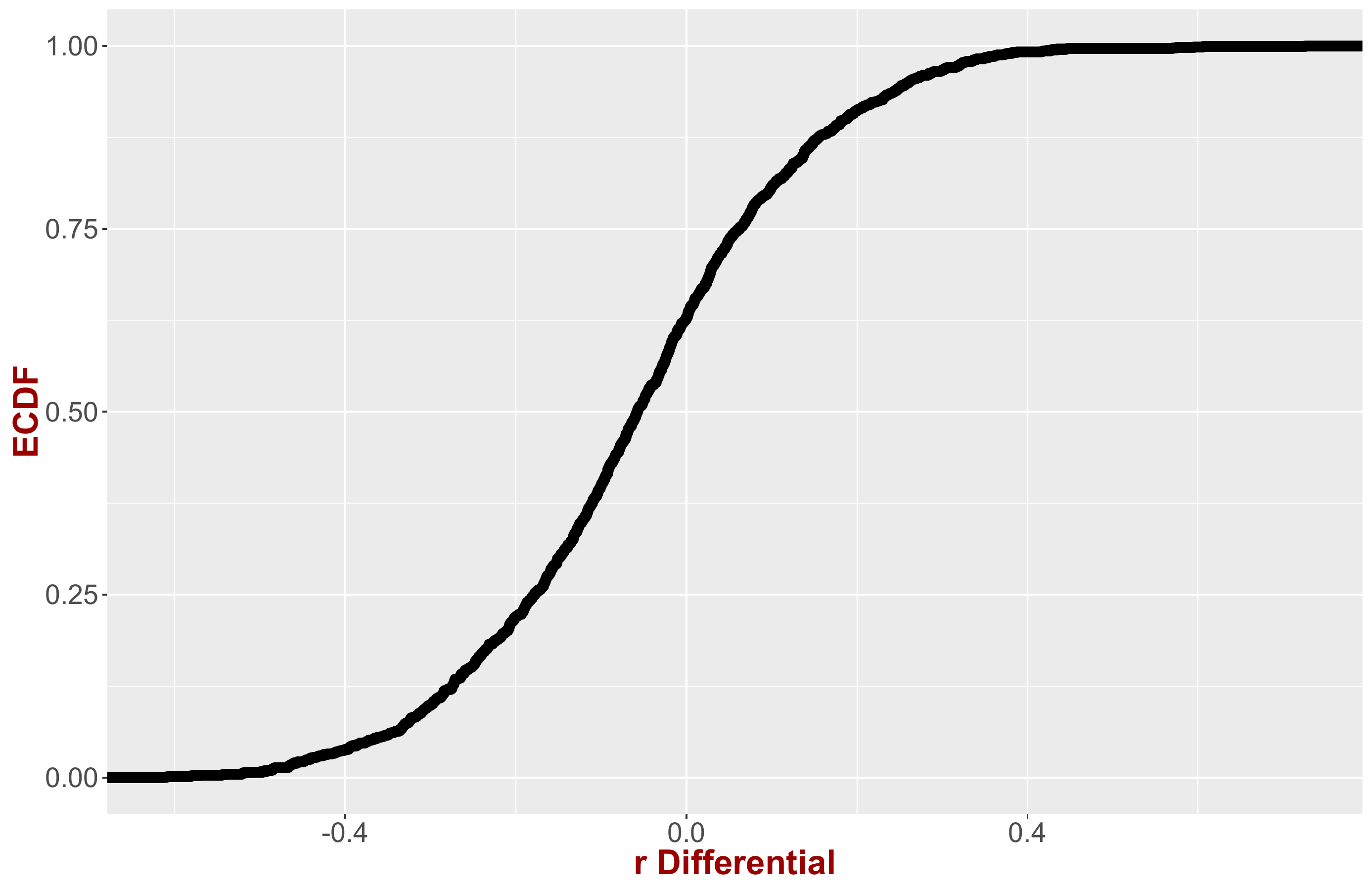}\includegraphics[scale=0.15]{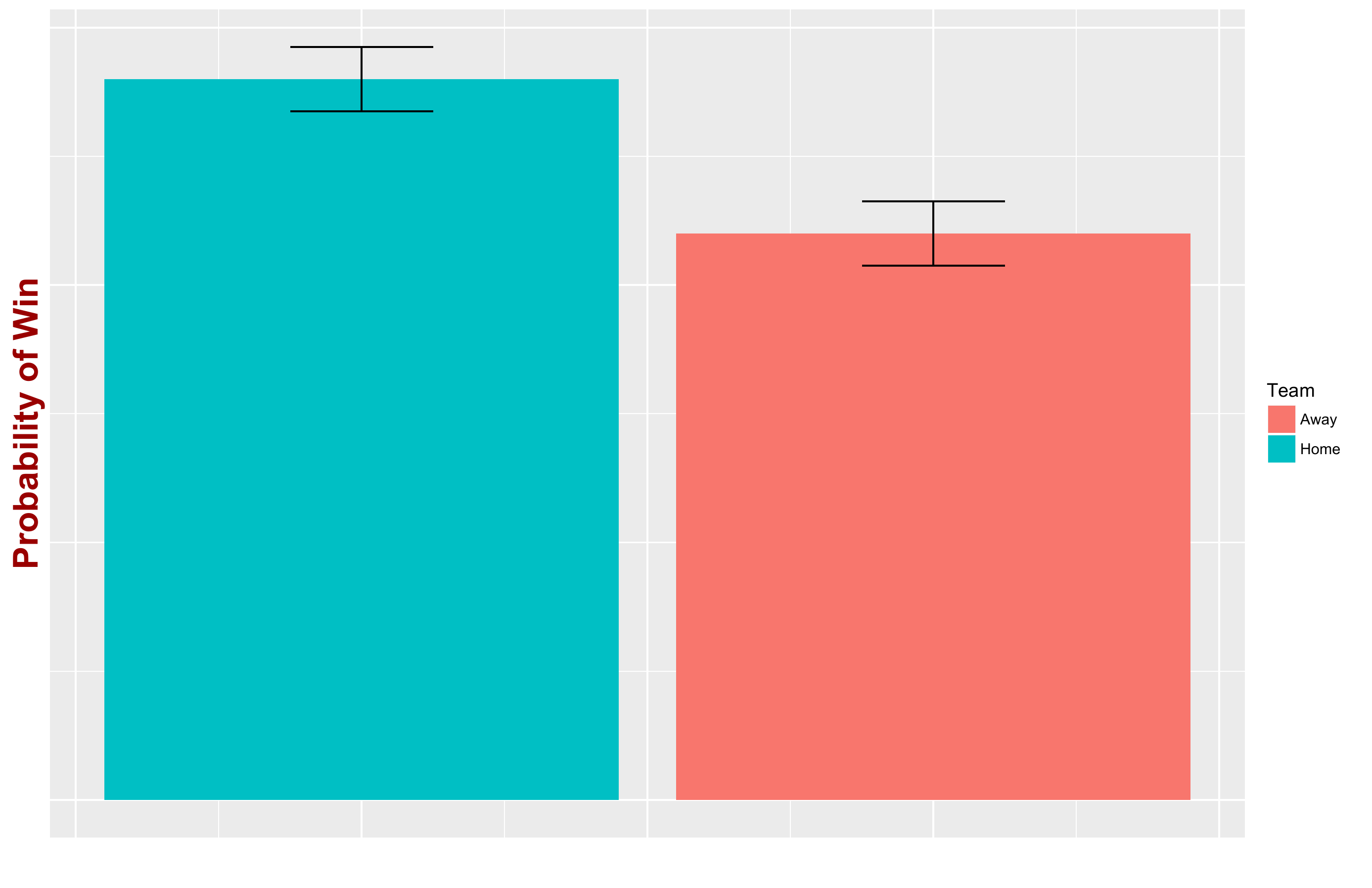}
\caption{{\bf Empirical cumulative distribution function for the paired differences of each feature}.  Based on the Kolmogorov-Smirnov test the features' ECDFs for the winning and losing teams are statistically different (at the significance level $\alpha=0.01$).  The probability mass function for the home team advantage is also presented.}
\label{fig:ecdf}
\end{figure}

Before delving into the details of the descriptive model, we perform some basic analysis that compares the game statistics and metrics used for obtaining the features we include in our regression model.  
In particular, given a game statistic $s_i$ (e.g., total offensive yards), we perform a paired comparison for this statistic between the winning and losing teams.  
In particular, for each continuous game statistic $s_i$ 
we compare the pairs ($s_{i,j}^+,~s_{i,j}^-$) with a paired t-test, where $s_{i,j}^+$ ($s_{i,j}^-$) is the value of $s_i$ for the winning (losing) team of the j$^{th}$ game in our dataset.  
Table \ref{tab:test} depicts the results of the two-sided paired t-tests for our continuous statistics together with the home team {\em advantage} observed in our data.    
As we can see all the differences are significantly different than zero (at the significance level of $\alpha=0.01$).  
Figure \ref{fig:ecdf} further presents the empirical cumulative distribution function (ECDF) for the paired differences for all the statistics as well as the probability mass function (PMF) for the distribution of the wins among home and visiting teams.  
For example, we can see that in only 20\% of the games the winning team had more turnovers as compared to the losing team.                        
We further perform the Kolmogorov-Smirnov test for the ECDFs of the considered statistics for the winning and losing teams.  
The tests reject the null hypothesis at the significance level of $\alpha=0.01$ for all cases, that is, the cumulative distribution of the features is statistically different for the winning and losing teams. 

\begin{table}[h]
  \begin{center} 
    \begin{tabular}{l|c}
    \toprule
			\bf Feature & \bf Average paired difference \\
      \midrule
Total Yards & 51.78 $^{***}$\\
Penalty Yards & -3.29$^{*}$\\
Turnovers & -1.04$^{***}$\\
Possession Time (sec) & 211.79$^{***}$ \\
$\ratio$ &  -0.06$^{***}$ \\
\bottomrule
\toprule
			{\em Home Team Advantage} & $56.03\%\pm 2.49\%$ \\
      \midrule
\bottomrule
    \end{tabular}
    \vspace{0.2cm}
       \caption{Paired t-test for the considered game statistics. The difference represents $\overline{s_{i,j}^+}-\overline{s_{i,j}^-}$. Significance codes:    *** : p $<$ .001, ** : p $<$ .01, * : p $<$ .05.  The home team advantage is also presented.}
    \label{tab:test}
  \end{center}
\end{table}


Our basic data analysis above indicates that the distribution of the statistics considered is significantly different for the winning and losing teams.  
However, we are interested in understanding which of them are good explanatory variables of the probability of winning a game.  
To further delve into the details, we use our data to train the Bradley-Terry regression model and we obtain the results presented in Table \ref{tab:regr}.  
Note here that, as it might be evident from the aforementioned discussion, we do not explicitly incorporate a feature for distinguishing between the home and the visiting team.  
Nevertheless, the response variable is the probability of the home team winning, while the features capture the differential of the respective statistics between the home and road team (i.e., the difference is ordered).  
Therefore, the intercept essentially captures the home team advantage - or lack thereof depending on the sign of the coefficient.   
In fact, setting all of the explanatory variables equal to zero provides us a response equal to $\Pr(\winrv_{ij}|\bf 0)$ $= 0.555$, which is equal to the home team advantage as discussed above.  
Furthermore, all of the coefficients - except the one for the possession time differential - are statistically significant.  
However, the impact of the various factors as captured by the magnitude of the coefficients range from weak to strong.  
For example, the number of total yards produced by the offense seem to have the weakest correlation with the probability of winning a game (i.e., {\em empty} yards).  
On the contrary committing turnovers quickly deteriorates the probability of winning the game and the same is true for an unbalanced offense.  
Finally, in S1 Text we present a standardized version of our model. 

\begin{table}[h]
  \begin{center}
    \begin{tabular}{l|c}
    \toprule
			\bf Feature & \bf Coefficient \\
      \midrule
Intercept &  0.22**\\
Total Yards differential & 0.01***\\
Penalty Yards differential & -0.02***\\
Turnovers differential & -1.05***\\
Possession Time differential & 0.0001 \\
$\ratio$ differential &  -3.18*** \\
$\Delta$ {\tt SportsNetRank} & 0.04*** \\
\bottomrule
    \end{tabular}
    \vspace{0.2cm}
       \caption{Coefficients of our Bradley-Terry regression model for the random variable $\winrv_{ij}$. Significance codes: *** : p $<$ .001, ** : p $<$ .01, * : p $<$ .05.}
    \label{tab:regr}
  \end{center}
\end{table}

While the direction of the effects for these variables are potentially intuitive for the coaching staff of NFL teams, the benefit of our quantifying approach is that it assigns specific magnitude to the importance of each factor.  
Clearly the conclusions drawn from  the regression cannot and should not be treated as causal.  
Nevertheless, they provide a good understanding on what is correlated with winning games.    
For example, if a team wins the turnover battle by 1 it can expect to obtain an approximately 20\% gain in the winning probability (all else being constant), while a 10-yard differential in the penalty yardage is correlated with just a 5\% difference in the winning probability.  
Hence, while almost all of the factors considered are statistically significant, some of them appear to be much more important as captured by the corresponding coefficients and potential parts of the game a team could work on.  
Again, this descriptive model {\bf does not provide a cause-effect relationship between the covariates considered and the probability of winning}.  

\begin{figure}[h]
\begin{center}
\vspace{-1in}
\includegraphics[angle=270,scale=0.5]{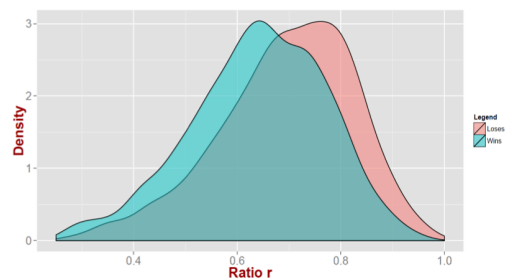}
\vspace{-1.3in}
\caption{{\bf Model validity}. Our model is trained within the range of input variable/statistics values on the left table.  The figure on the right presents the probability density function for $\ratio$ for the winning and losing instances respectively. }
\label{fig:quant-ratio}
\end{center}
\end{figure}

\begin{figure}[!h]
\begin{center}
\includegraphics[scale=0.5]{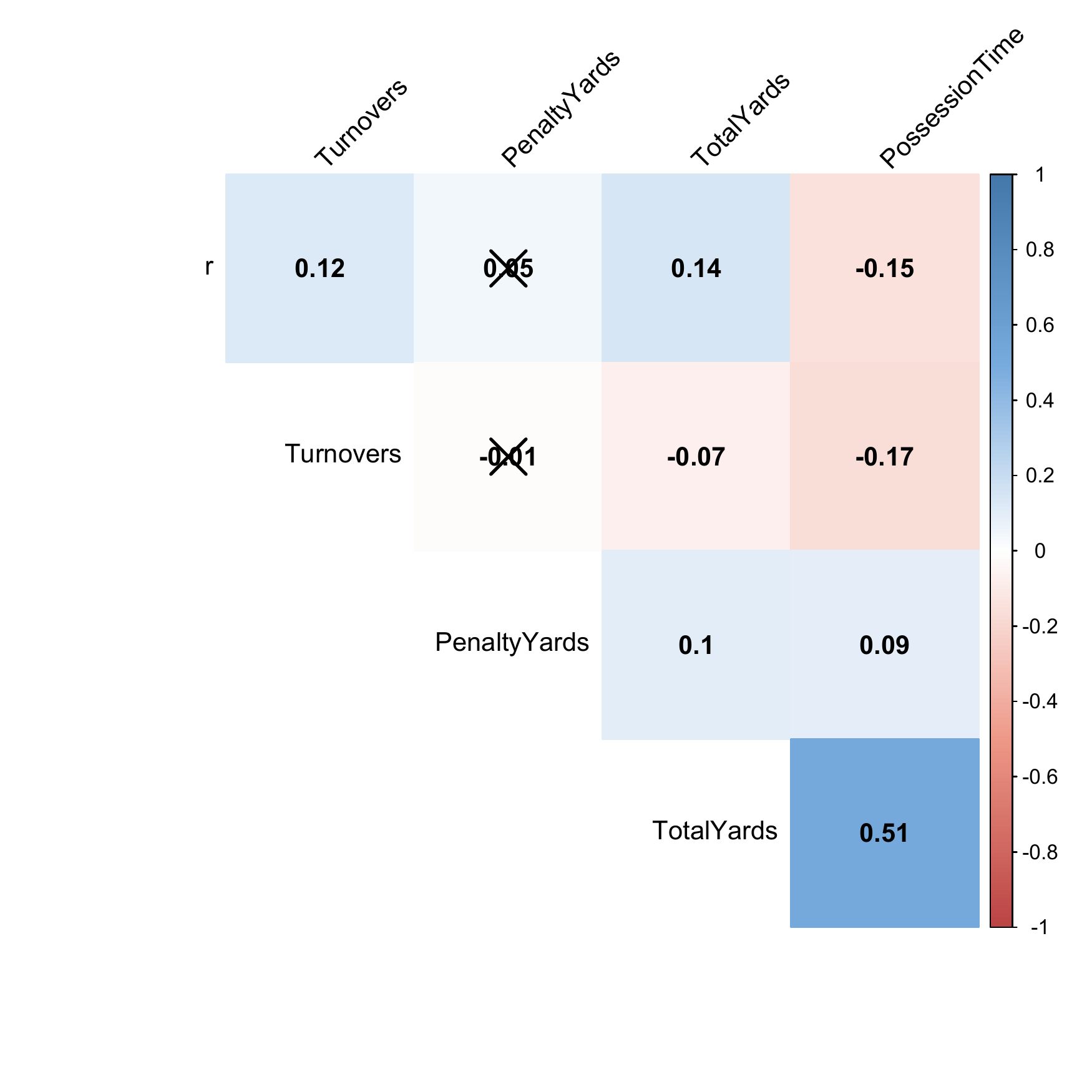}
\caption{{\bf Correlograms}. Correlations between the different variables considered for obtaining the features for {\method}.  Insignificant correlations  are crossed out.}
\label{fig:corr}
\end{center}
\end{figure}

Before turning to the {\method} predictive engine we would like to further emphasize and reflect on how one should interpret and use these results.  
For example, one could be tempted to focus on the feature with the coefficient that exhibits the maximum absolute magnitude, that is, ratio $\ratio$, and conclude that making $\ratio = 0$, i.e., calling only run plays will increase the probability of winning, since the negative differential with the opposing team will be maximized.  
However, this is clearly not true as every person with basic familiarity with American football knows.  
At the same time the regression model is not contradicting itself.  
What happens is that the model developed - similar to any data driven model - is {\em valid} only for the range of values that the input variables cover.  
Outside of this range, the generalized linear trend might still hold or not.  
For example, Figure \ref{fig:quant-ratio} depicts the distribution of ratio $\ratio$ for the winning and losing teams.  
As we can see our data cover approximately the range $\ratio\in[0.3, 0.98]$ and the trend should only be considered valid within this range (and potentially within a small $\epsilon$ outside of this range).  
It is interesting also to observe that the mass of the distribution for the winning teams is concentrated around $\ratio \approx 0.64$, while it is larger for the losing teams ($\ratio \approx 0.8$).    
We also present at the same figure a table with the range that our features cover for both winning and losing teams.   
Furthermore, to reiterate, the regression model captures merely correlations (rather than cause-effect relations). 
Given that some of the statistics involved in the features are also correlated themselves (see Figure \ref{fig:corr}) and/or are result of situational football, makes it even harder to identify real causes.  
For instance, there appears to be a small but statistically significant negative correlation between ratio $\ratio$ and possession time.  
Furthermore, a typical tactic followed by teams leading in a game towards the end of the fourth quarter is to run the clock out by calling running plays.  
This can lead to a problem of reverse causality; a reduced ratio $\ratio$ for the leading team as compared to the counterfactual $\ratio$ expected had the team continued its original game-plan, which can artificially deflate the actual contribution of $\ratio$ differential on the probability of winning.  
Similarly, teams that are trailing in the score towards the end of the game will typically call plays involving long passes in order to cover more yardage faster.  
However, these plays are also more risky and will lead to turnovers more often, therefore, inflating the turnover differential feature.  
Nevertheless, this is always a problem when a field experiment cannot be designed and only observational data are available.  
While we cannot claim causal links between the covariates and the output variable, in what follows we present evidence that can {\em eliminate} the presence of reverse causality for the scenarios described above. 

{\bf Reverse Causality: }
In what follows we examine the potential for reverse causality.  
To fast forward to our results, we do not find strong evidence for it.  
To reiterate, one of the problems with any model based on observational data is the direction of the effects captured by the model.  
For example, in our case 
teams that are ahead in the score towards the end of the game follow a ``conservative'' play call, that is, running the football more in order to minimize the probability of a turnover and more importantly use up valuable time on the clock.  
Hence, this can lead to a decreasing ratio $\ratio$.  
Therefore, the negative coefficient for the $\ratio$ differential in our regression model might be capturing reverse causality/causation. 
Winning teams artificially decrease $\ratio$ due to conservative play calling at the end of the game.  
Similarly, teams that are behind in score towards the end of the game follow a more ``risky'' game plan and hence, this might lead to more turnovers (as compared to the other way around).  

\begin{figure}[h]
\begin{center}
\includegraphics[scale=0.25]{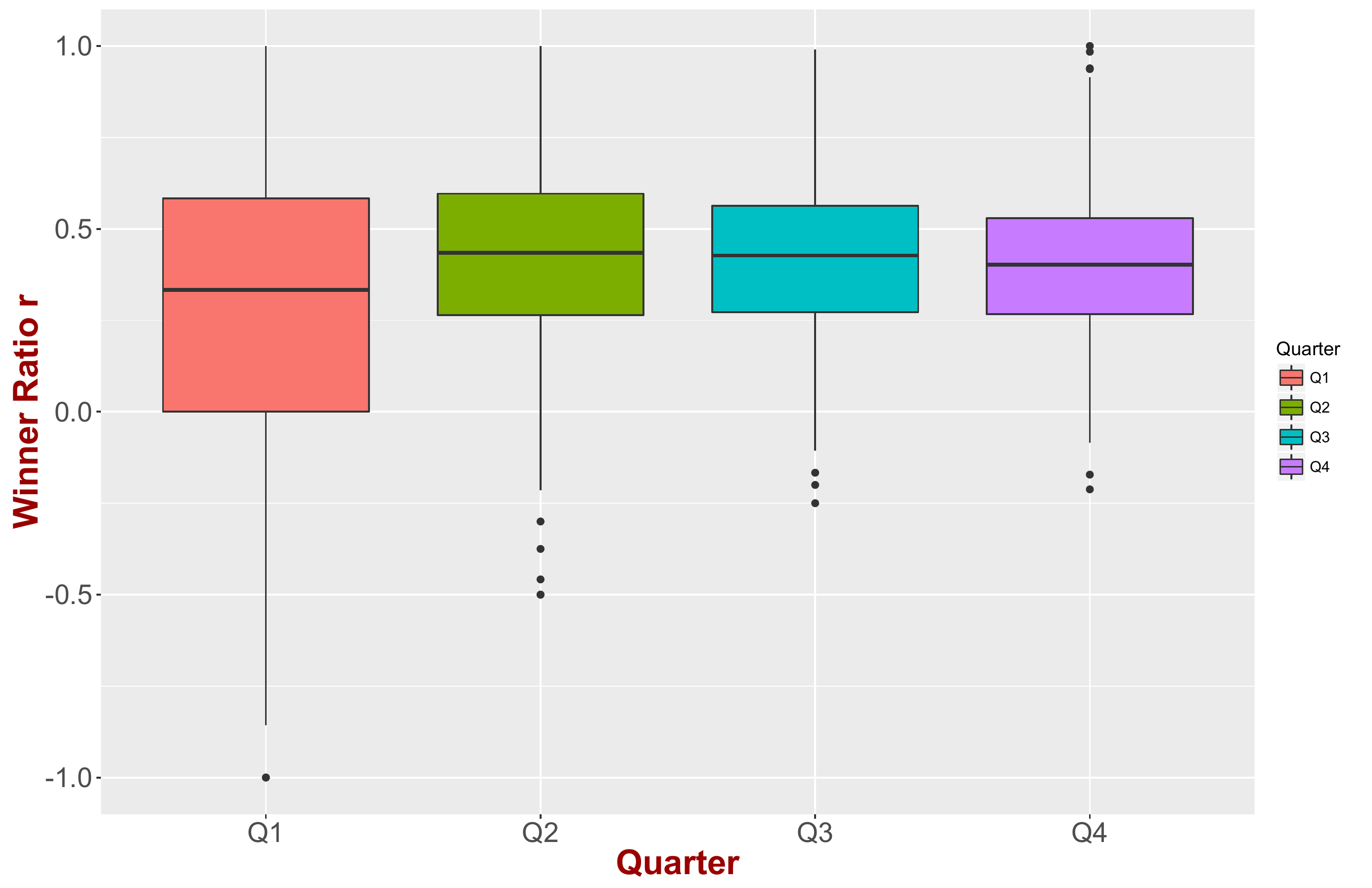}\includegraphics[scale=0.25]{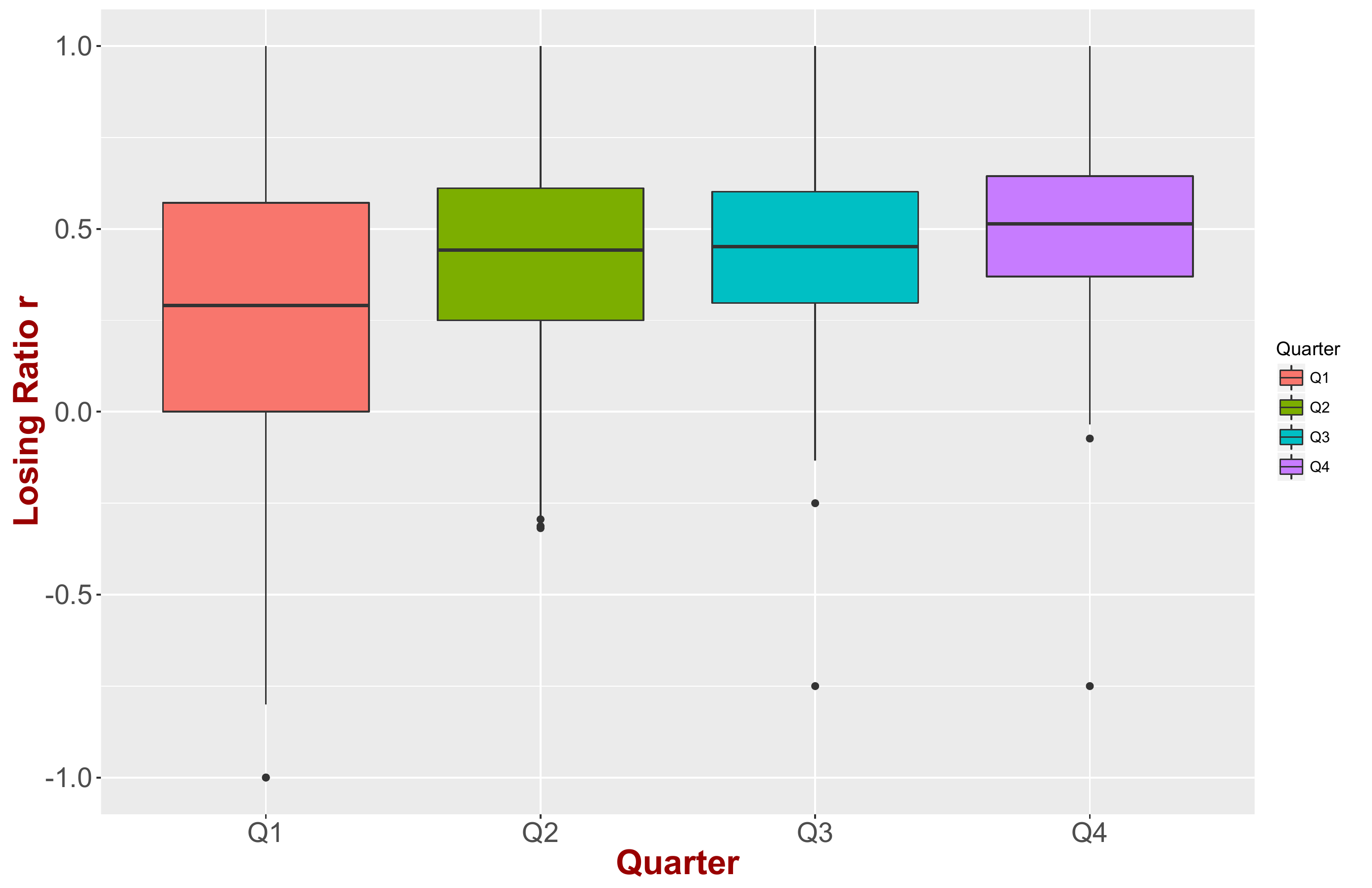}
\caption{{\bf Evolution of $\ratio$ through the game}. Ratio $\ratio$ is stable after the first quarter for both winning (left figure) and losing (right figure) teams, allowing us to reject the reverse causation hypothesis for $\ratio$.}
\label{fig:r_wins}
\end{center}
\end{figure}

One possible way to explore whether this is the case is to examine how the values of these two statistics change over the course of the game.  
We begin with ratio $\ratio$.  
If the reverse causation hypothesis were true, then the ratio $\ratio$ for the winning team of a game would have to reduce over the course of the game.  
In order to examine this hypothesis, we compute the ratio $\ratio$ at the end of each quarter for both the winning and losing teams.  Figure \ref{fig:r_wins} presents the results.  
As we can see during the first quarter there is a large variability for the value of $\ratio$ as one might have expected mainly due to the small number of drives.  
However, after the first quarter it seems that the value of $\ratio$ is stabilized.  
There is a slight decrease (increase) for the winning (losing) team during the fourth quarter but this change is not statistically significant.  
Therefore, we can more confidently reject the existence of reverse causality for ratio $\ratio$.

\begin{figure}[h]
\begin{center}
\includegraphics[width=3in,height=3in]{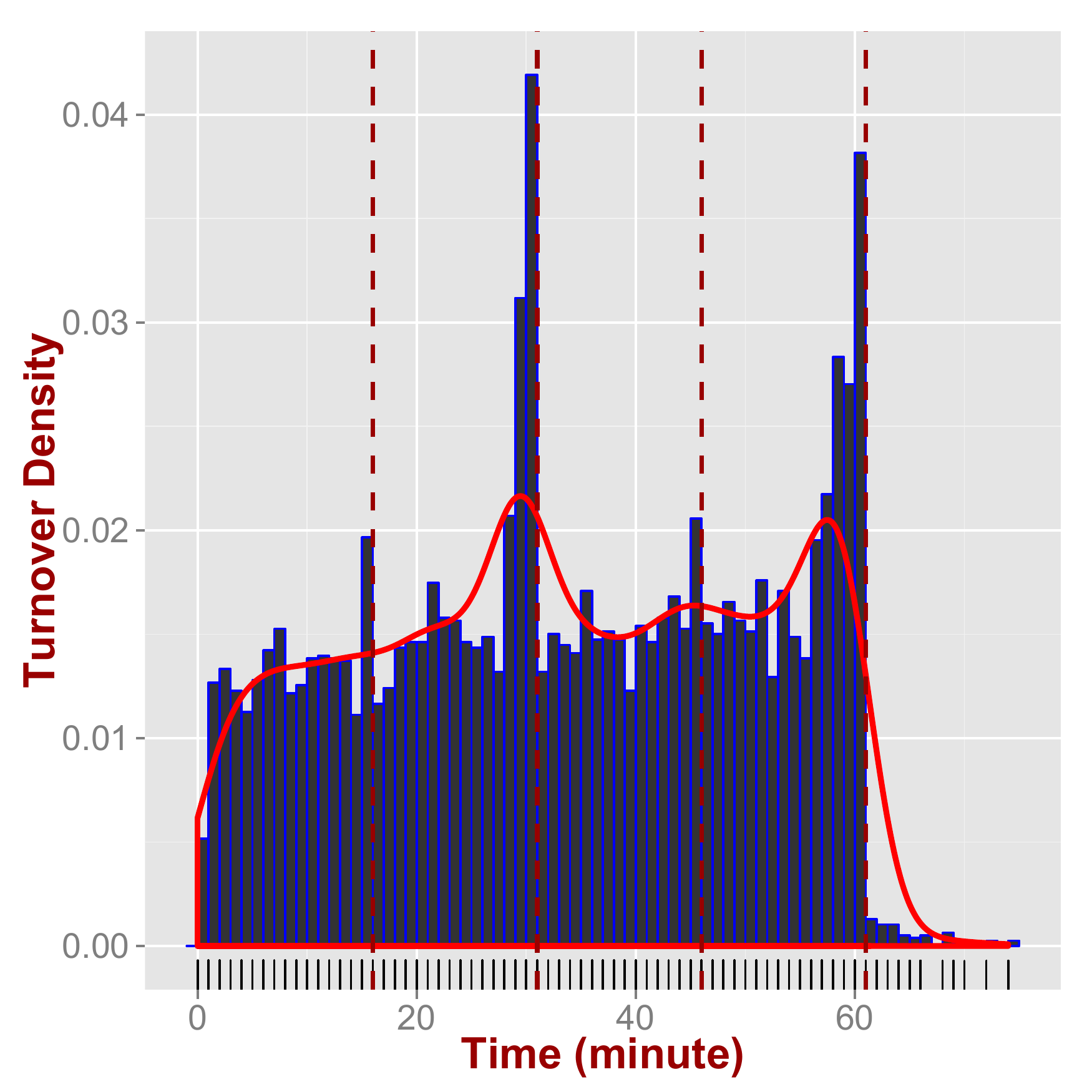}
\caption{{\bf Temporal dynamics of turnovers}. Turnovers spike towards the end of each quarter, with the highest density appearing during the two-minute warning.}
\label{fig:turnovers-time}
\end{center}
\end{figure}

We now focus our attention on the turnovers and the potential reverse causation with respect to this feature.  
In order to examine this hypothesis, we obtain from our data the time within the game (at the minute granularity) that turnovers were committed by the winning and losing teams.  
We then compare the paired difference for the turnover differential until the end of the third quarter for each game.  
Our results show that the winning teams commit fewer turnovers than their losing opponents by the end of the third quarter ($p$-value $<$ 0.01), further supporting that avoiding turnovers will ultimately lead to a win.  
Of course, as we can see from Figure \ref{fig:turnovers-time}, there is a spike of turnovers towards the end of each half (and smaller spikes towards the end of each quarter).  
These spikes can be potentially explained from the urgency to score since either the drive will stop if the half ends or the game will be over respectively.  
However, regardless of the exact reasons for these spikes, the main point is that by committing turnovers, either early in the game (e.g., the first three quarters) or late, the chances of winning the game are significantly reduced.  

In conclusion, our model provides quantifiable and actionable insights but they need to be carefully interpreted when designing play actions based on it.

\subsection{{\method} Prediction Engine}
\label{sec:prediction}

We now turn our attention on how we can use the above model in order to predict the outcome of a future game.  
In a realistic setting, in order to be able to apply this regression model we will need to provide as an input the team statistics/features.   
This is by itself a separate prediction problem, namely, a team performance prediction problem.  
Hence, we begin by evaluating the prediction performance of the Bradley-Terry regression model itself using traditional machine learning evaluation methods.  
In particular, we evaluate the prediction accuracy of our model through cross validation.  
In this way we do not need to predict the value of the features but we explore the accuracy of the pure regression model.   
Using 10-fold cross validation we obtain an accuracy of {\bf 84.03\%} $\pm$ {\bf 0.35\%}.  
To reiterate this performance is conditional to the input features being known. 
From the inputs required for our model only two are known before the matchup, namely, the home team (which will allow us to formulate the response variable and the rest of the features appropriately) and the {\tt SportsNetRank} differential. 
Thus, how can we predict the rest of the features, since in a realistic setting we will not know the performance of each team beforehand?  
Simply put, our {\method} prediction engine will need to first estimate the two teams statistics/features (i.e., total yards, penalty yards, etc.) and then use the Bradley-Terry regression model to predict the winning team.  

\begin{figure}[h]
\begin{center}
\includegraphics[scale=0.8]{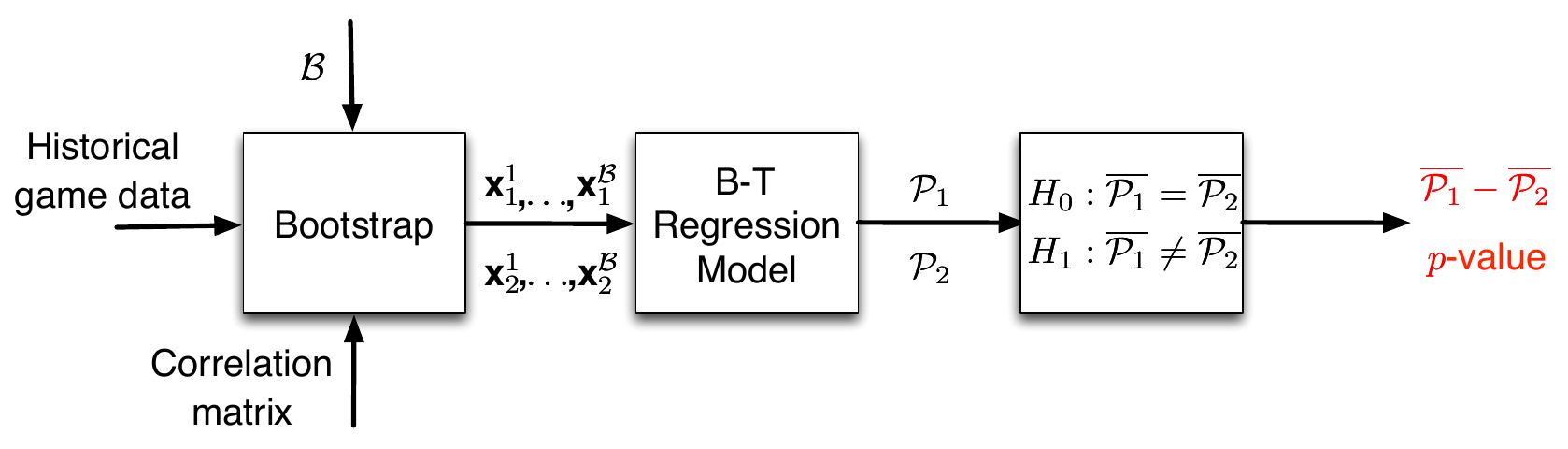}
\caption{{\bf Football Matchup Prediction ({\method})}. The proposed prediction engine consists of 3 modules; a bootstrap module, a regression module and a statistical test module.  }
\label{fig:system}
\end{center}
\end{figure}

The most straightforward way for this task is to use historic game data from the current season and calculate descriptive statistics such as the mean or the median of each performance indicator of the teams and then compute the model's features.  
The problem with this approach is that using a measure of central tendency does not accurately capture the variability in the teams' performance.  
Therefore, we propose to utilize statistical bootstrap in order to resample with replacement $\boots$ times the empirical distribution for each one of the team statistics from the observed sample.  
Multiple draws from the historic data will allow to properly characterize the input features of the model. 
For every resampling we can calculate the win probability for the home team using the Bradley-Terry regression model and ultimate obtain 
a confidence interval for the win probability for each of the teams.  
This essentially allows us to statistically compare the chances of each team winning the game.  
Figure \ref{fig:system} illustrates the components of {\method}.  

The first part of {\method} obtains three inputs; (i) historical information for games of the current season, (ii) the number of bootstrap samples $\boots$ to obtain and (iii) the correlation matrix between the features to be resampled.  
In particular, for every team $T$ we have a matrix $\mathbf{M}_T$ each row of which represents a game in the current season, while the columns correspond to the five different statistics used in the features of the Bradley-Terry model.  
In the case of ``simple'' bootstrapping we would uniformly at random select for each performance statistic one row (i.e., one of the past performances of the team with respect to this feature) and hence, we would obtain a resampled vector $\mathbf{x}_T^i$ that represents a potential performance for $T$ given its past.  
However, there are two factors that we need to take into consideration.  
First, more recent games might be more representative of recent adjustments (or roster losses due to injuries) as compared to performances during the first weeks of the season.  
In order to control for this, instead of sampling the rows of $\mathbf{M}_T$ for every feature uniformly at random, we bias the sampling probabilities to favor the last $k$ games of the team.  
Second, sampling the performance statistics independently can lead to vectors $\mathbf{x}_T^i$ that do not exhibit the correlations that are present in the actual data.  
To reiterate Figure \ref{fig:corr} represents the correlations between the different pairs of performance statistics.  
For example, as we can see the total yards and the possession time exhibit a medium to strong level of correlation.  
This means that when we sample for the total yards of the i$^{th}$ bootstrap sample, we should not sample the possession time independently, but rather select the possession time from the same game/row of $\mathbf{M}_T$.  
This essentially mimics the block bootstrapping approach \cite{kunsch89} used for time-series data to keep the dependencies between consecutive time-points.  
The rest of the correlations between the features are fairly weak (and some also insignificant) and hence, we proceed as normal with the rest of the statistics.  

Once bootstrapping is completed its output is essentially a set of potential future performances for each team as captured through the obtained vectors.  
Simply put for each of the two competing teams we have bootstrapped vectors $\mathbf{x}_1^1, \mathbf{x}_1^2,\dots,\mathbf{x}_1^{\boots}$ and $\mathbf{x}_2^1, \mathbf{x}_2^2,\dots,\mathbf{x}_2^{\boots}$ respectively that capture the predicted game stats for the home and visiting team respectively.  
These vectors form the input for our regression model - in fact, the input for our model is $\inputvec^j = \mathbf{x}_1^j-\mathbf{x}_2^1$ - which provides a set of winning probabilities for each team, i.e., $P_1 = \{\Pr(\winrv_{12}|\inputvec^1)\dots,\Pr(\winrv_{12}|\inputvec^{\boots})\},~P_2 = \{1-P_1^1,\dots,1-P_1^{\boots}\}$. 
Once we obtain these probability sets, we finally perform a hypothesis test to identify whether the two sets represent probabilities that are statistically different at a predefined significance level $\alpha$: 

\vspace{-0.15in}
\begin{eqnarray}
H_0: & \overline{P_1} = \overline{P_2} \label{eq:null}\\
H_1: & \overline{P_1} \neq \overline{P_2} \label{eq:alt}
\end{eqnarray}

If the null hypothesis is rejected, then the sign of the difference $\overline{P_1}-\overline{P_2}$ will inform us about the team that is most probable to win the matchup.  
If the null hypothesis is not rejected, then we can predict a tie. 
It should be evident that our predictive engine cannot be applied during the first week of the season, while the weekly variability of teams' performance can be fully exploited in later parts in the season.  
Considering that, we set $k=5$ in the current version of {\method} and for each NFL season we start our predictions from Week 6.  
When focusing on a specific NFL season for predicting the game outcomes, we train our model using data from the rest of the seasons.  
Note here that the coefficients presented in Table \ref{tab:regr} are obtained using all 7 years worth of data.  
When training the model using all the possible subsets of 6 seasons the obtained coefficients differ but not in any meaningful way. 
We further set $\boots = 1,000$ and $\alpha  = 0.05$.  
The overall accuracy of {\method} is {\bf 63.4\%} with a {\bf standard error of 1.3\%}. 
This performance is no statistically different than the accuracy of the current state-of-the-art NFL prediction systems.  
For example, Microsoft's Cortana system exhibited an accuracy of approximately 64.5\% \cite{cortana} during the last two seasons that it has been operating.  
Similarly, the prediction accuracy of ESPN's FPI is 63\% as well \cite{espn-fpi}.
Furthermore, we randomly sampled forecasts of sports analysts from major networks (ESPN, NFL network, CBS and FOX sports) \cite{nfl-expert}. 
Our predictions were on average better than approximately 60\% of the expert predictions.  

Delving more into the evaluation of our predictive engine we present the accuracy for each season in Table \ref{tab:results_accuracy}.  
We also provide the accuracy of a baseline system, where the winner of a game is predicted to be the team with the better running win-loss percentage through the current week.  
If two teams have the same win-loss percentage the home team is chosen as the winner since there is a slight winning bias for the home team as we have seen earlier.   
Note here that the baseline is very similar to the way that the league ranks the teams and decides on who will qualify for the playoffs (excluding our tie-breaker process and the league's rules with respect to the division).  
As we can see our predictive engine improves over the baseline by approximately 9\%.  

\begin{table}[ht]
\centering
 \begin{tabular}{c|lcc}
    \toprule
    {\bf Year} & {\bf Regression} & {\bf Baseline}\\
    \midrule
2009 & 0.66 & 0.57 \\ 
\hline
2010 & 0.60 & 0.5 \\ 
\hline 
2011 & 0.68 & 0.58 \\ 
\hline
2012 & 0.72 & 0.56\\ 
\hline 
2013 & 0.55 & 0.5 \\ 
\hline
2014 & 0.66 & 0.56\\ 
\hline
2015 & 0.57 & 0.53 \\ 
 \bottomrule
\end{tabular}
\caption{{\bf Prediction accuracy}. {\method} outperforms the baseline prediction based on win-loss standings every season in our dataset.  The overall accuracy of our system is 63.4\%.}
\label{tab:results_accuracy}
\end{table}

\begin{figure}[h]
\begin{center}
\includegraphics[scale=0.5]{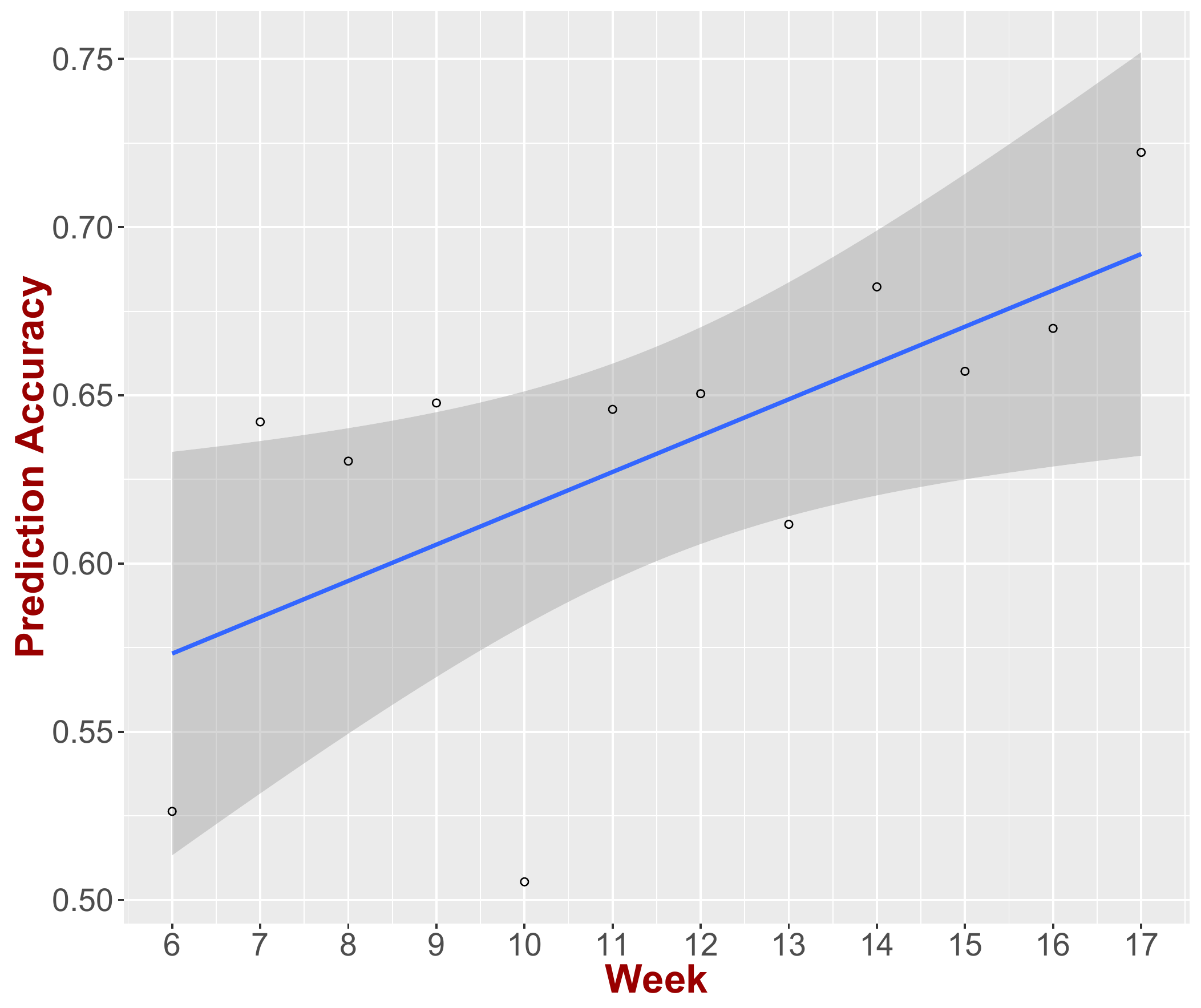}
\caption{{\bf Prediction accuracy VS week}. During the last part of the season the bootstrap engine can exploit the variability of a team's performance better, hence, providing better prediction accuracy. The linear trend slope is 0.01 (p-value$<$0.05, $R^2 = 0.41$).}
\label{fig:accuracy_week}
\end{center}
\end{figure}

One of the reasons we utilize bootstrap in our prediction system is to better capture the variability of the teams' performances.  
As one might expect this variability is better revealed as the season progresses.  
During a stretch of few games it is highly probable to have a team over/under-perform \cite{gilovich1985hot}.  
Hence, the bootstrap module during the beginning of the season might not perform as accurately as during the end of the season.  
In order to examine this we calculate the accuracy of our prediction system focusing on games that took place during specific weeks in every season.  
Figure \ref{fig:accuracy_week} presents our results, where we see that there is an increasing trend as the season progresses.    

\begin{figure}[h]
\begin{center}
\includegraphics[scale=0.5]{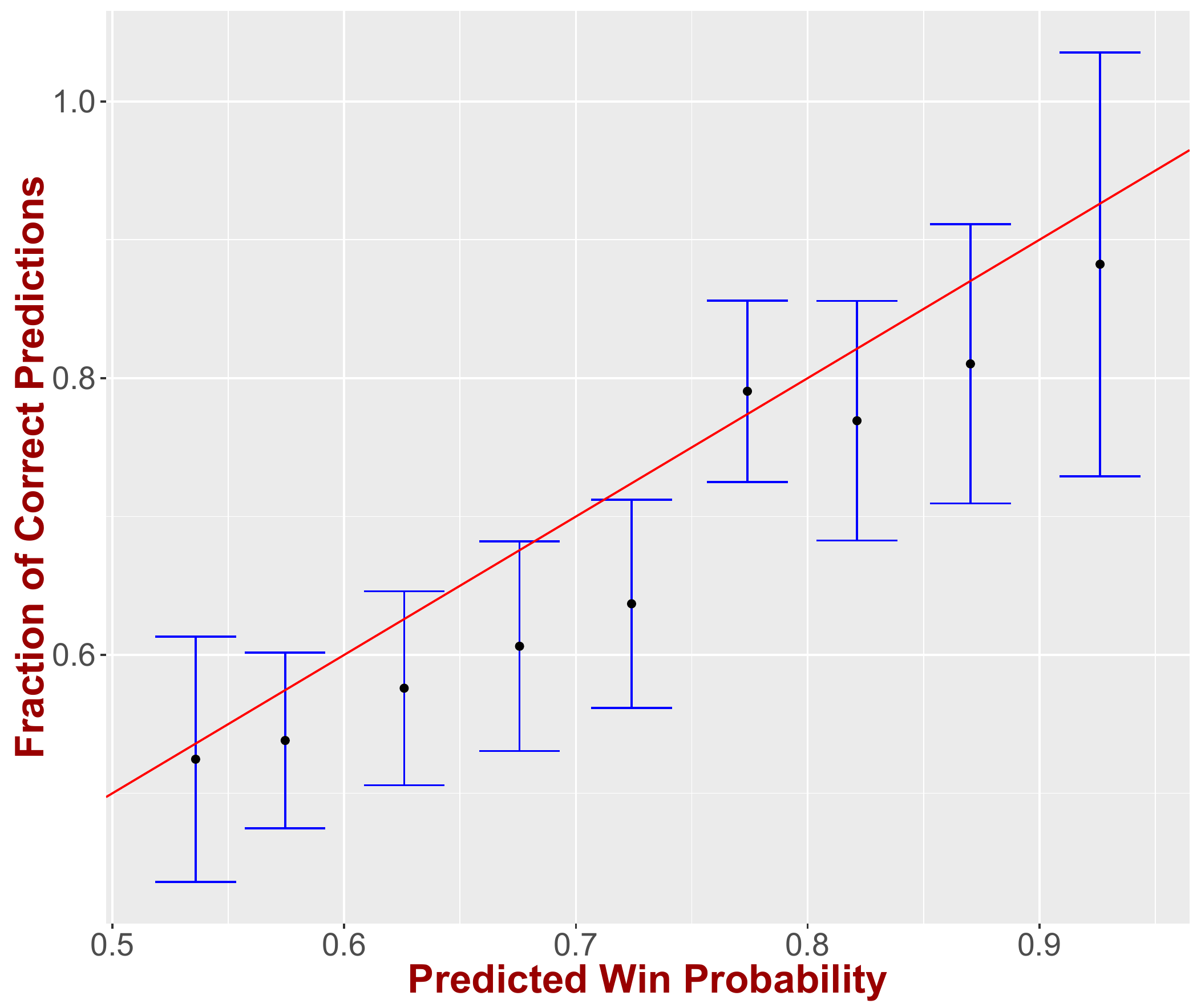}
\caption{{\bf Probability Accuracy}. The win probability provided by our model is in alignment with the fraction of the games won by the favorite for the corresponding win probability. }
\label{fig:prob_accuracy}
\end{center}
\end{figure}

Finally, we examine the accuracy of {\method}'s predicted probabilities.  
In order to evaluate this we would ideally want to have the game played several times.  
If the favorite team were given a 75\% probability of winning, then if the game was played 100 times we would expect the favorite to win 75 of them.  
However, we cannot have the game play out more than once and hence in order to evaluate the accuracy of the probabilities we will use all the games in our dataset.  
In particular, if the predicted probabilities were accurate, when considering all the games where the favorite was predicted to win with a probability of $x\%$, then in $x\%$ of these games the favorite should have won. 
Given the continuous nature of the probabilities we quantize them into groups that cover a 5\% probability range (with only exception being the range [90\%,100\%), since there are very few games in the corresponding sub-groups).  
Figure \ref{fig:prob_accuracy} presents on the y-axis the fraction of games where the predicted favorite team won, while the x-axis corresponds to the predicted probability of win.  
As we can see the data points - when considering their 95\% confidence intervals - fall on the $y=x$ axis, which translates to a very accurate probability inference.  
The corresponding linear regression provides a slope with a 95\% confidence interval of [0.76, 1.16] ($R^2 = 0.94$), which essentially means that we cannot reject the null hypothesis that our data fall on the line $y=x$ where the slope is equal to 1. 

\section{Discussion}
\label{sec:discussion}

In this study we have used data from the last 7 NFL seasons to first examine whether coaches are making rational, informed by data, decisions.  
Our results indicate that this is not the case after examining two specific play calls that teams face very often in each game, PAT and fourth down.  
Similar findings have been reported in the past \cite{doi:10.1080/00031305.1967.10479847,sackrowitz2000refining,RePEc:ucp:jpolec:v:114:y:2006:i:2:p:340-365} but one might have expected that with the recent ability of teams to collect and analyze much more detailed data for the game their playbooks would have been more data-driven.    
After rejecting the hypothesis of {\tt rational coaching}, we focus on identifying factors that can impact the probability of winning a game.  
We then combine this model with statistical bootstrap in order to predict the outcome of upcoming match ups.  

We would like here to emphasize on the fact that our analysis for the {\tt rational coaching} hypothesis, simply rejects the hypothesis.  
More specifically, we do not claim that it is always beneficial to try a two-point conversion.  
There are case where an extra point kick is the optimal case.  
For instance, if a team is down by 6 points and scores a touchdown to tie the game with 1 second left on the clock, it is obvious that the optimal strategy is to go for an extra point kick since the probability of success (approximately 95\%) is much higher as compared to converting the two points attempt (a little over 50\%).  
This is exactly what we also explained in Figure \ref{fig:point_ben} with respect to the fourth down decision.  
It is not always beneficial (i.e., $\mathbf{E}[P] > 0$) to go for it on fourth down but it is beneficial way more often than what coaches decide to go for it; an observation that allows us to reject the {\tt rational coaching} hypothesis! 
Moreover it might be the case that many of the coaching decisions are based on minimizing the variance of the outcome - and in this case the choice for the extra kick is rational - or the risk adjusted return (i.e., the expected gain divided by the variance).  
 
Our prediction results indicate that as the season progresses the performance of our system is improved; there is more information about the historic performance of a team that bootstrap can exploit.   
Earlier in the season this information is more sparse and our system's accuracy is expected to exhibit high variance.  
With respect to the performance of our prediction engine we would like to emphasize on the fact that it is very impressive if one considers that we do not account for game-specific information (e.g., injured players, teams resting players during the last week, etc.).  
Even without considering such information the system performs on average better than the experts, who are taking into consideration these signals, 60\% of the time.  
Hence, our system in its current form can be considered as providing an {\em anchoring} probability \cite{kahneman1973psychology} from which an expert can adjust his/her prediction using game-specific information.
  
Moreover, there is room for improving this anchoring probability further.  
In particular, one of the limitations of our system is that currently it does not incorporate any information for the schedule strength of teams.  
Simply put, when performing the bootstrap for the future performance of team $T$ we just extrapolate from the older performance of $T$ without considering the strength of the teams that it has played against.  
Teams might have put up a lot of offensive yards just because they have played with teams with poor defense. 
For example, in the 2014 NFL season the winner of NFC South (Carolina Panthers) had a season record less than 0.5.  
This means that Carolina had faced mainly teams with a losing record and hence, the corresponding team statistics might have been inflated as a result of the bootstrap process described above.  
However, our engine is easily adoptable to account for this.  
In particular, let us consider that we want to estimate the performance of team $T_1$ against $T_2$.  
In order to account for the strength of $T_2$, we can bias the re-sampling probabilities based on the propensity score \cite{austin2011introduction} of the set of teams $\mathcal{T}$ that $T_1$ has faced in the past. 
The propensity score is mainly used as a quasi-experimental technique, in order to match a treated sample with an untreated observational set based on a set of observable cofounders.  
In our setting, the propensity score $\pi_i$ for $T_i \in \mathcal{T}$ will essentially provide us with a ``similarity'' measure of $T_i$ with $T_2$.  
Consequently, the resampling probability for $T_i$ will be proportional to $\pi_i$. 
This biased resampling will allow us to incorporate/simulate in the prediction engine the potential interactions between the teams of the upcoming match ups.  
Including defense-oriented features in the propensity score matching will allow us to perform a more balanced prediction - i.e., consider both offense and defense - since defensive attributes are currently underrepresented in our prediction logistic regression model.  
This can potentially significantly improve the prediction performance of our engine.  


\section*{Supporting Information}

\noindent{\bf S1 Text. Standardized {\method}.}
\label{si:standardized_model}  
For easier interpretation and comparison between the different covariates of the model we present a standardized version of {\method}. 
In particular, for each feature of the model we subtract the corresponding mean and divide with the standard deviation.  
In this case, all the coefficients correspond to an one standard deviation change in the covariate and hence, a straight-forward comparison of the magnitude of the coefficient across covariates is possible.

\begin{table}[h]
  \begin{center}
    \begin{tabular}{l|c}
    \toprule
			\bf Feature & \bf Coefficient \\
      \midrule
Intercept &  0.50**\\
Total Yards differential & 1.82***\\
Penalty Yards differential & -0.83***\\
Turnovers differential & -2.08***\\
Possession Time differential & -0.07 \\
$\ratio$ differential &  -0.63*** \\
$\Delta$ {\tt SportsNetRank} & 0.55*** \\
\bottomrule	
    \end{tabular}
    \vspace{0.2cm}
       \caption{Standardized coefficients of our Bradley-Terry regression model for the random variable $\winrv_{ij}$. Significance codes: *** : p $<$ .001, ** : p $<$ .01, * : p $<$ .05.}
    \label{tab:regr}
  \end{center}
\end{table}

\small
\bibliography{ta}

\begin{thebibliography}{10}
\providecommand{\url}[1]{\texttt{#1}}
\providecommand{\urlprefix}{URL }
\expandafter\ifx\csname urlstyle\endcsname\relax
  \providecommand{\doi}[1]{doi:\discretionary{}{}{}#1}\else
  \providecommand{\doi}{doi:\discretionary{}{}{}\begingroup
  \urlstyle{rm}\Url}\fi
\providecommand{\bibAnnoteFile}[1]{%
  \IfFileExists{#1}{\begin{quotation}\noindent\textsc{Key:} #1\\
  \textsc{Annotation:}\ \input{#1}\end{quotation}}{}}
\providecommand{\bibAnnote}[2]{%
  \begin{quotation}\noindent\textsc{Key:} #1\\
  \textsc{Annotation:}\ #2\end{quotation}}
\providecommand{\eprint}[2][]{\url{#2}}

\bibitem{lamb12}
Lamb C, Hair J, McDaniel C (2012) Essentials of Marketing.
\newblock ISBN-13: 978-0538478342.
\bibAnnoteFile{lamb12}

\bibitem{economist-superbowlxlix}
Economist T (2015).
\newblock Game theory in american football: Defending the indefensible.
\newblock
  \url{http://www.economist.com/blogs/gametheory/2015/02/game-theory-american-football}.
\newblock Accessed: 2016-01-12.
\bibAnnoteFile{economist-superbowlxlix}

\bibitem{clark2013going}
Clark T, Johnson A, Stimpson A (2013) Going for three: Predicting the
  likelihood of field goal success with logistic regression.
\newblock In: The 7th Annual MIT Sloan Sports Analytics Conference.
\bibAnnoteFile{clark2013going}

\bibitem{Pfitzner14}
Pfitzner C, Lang S, Rishel T (2014) Factors affecting scoring in nfl games and
  beating the over/under line.
\newblock The Sport Journal .
\bibAnnoteFile{Pfitzner14}

\bibitem{warner10}
Warner J (2010) Predicting margin of victory in nfl games: Machine learning vs.
  the las vegas line.
\newblock Technical Report .
\bibAnnoteFile{warner10}

\bibitem{stair08}
Stair A, Day A, Mizak D, Neral J (2008) The factors affecting team performance
  in the nfl: does off-field conduct matter?
\newblock Economics Bulletin 26: 1--9.
\bibAnnoteFile{stair08}

\bibitem{fewell2012basketball}
Fewell J, Armbruster D, Ingraham J, Petersen A, Waters J (2012) Basketball
  teams as strategic networks.
\newblock PLOS ONE 101371/journalpone0047445 .
\bibAnnoteFile{fewell2012basketball}

\bibitem{Vilar2013}
Vilar L, Ara{\'u}jo D, Davids K, Bar-Yam Y (2013) Science of winning soccer:
  Emergent pattern-forming dynamics in association football.
\newblock Journal of Systems Science and Complexity 26: 73--84.
\bibAnnoteFile{Vilar2013}

\bibitem{bar2007action}
Bar-Eli M, Azar OH, Ritov I, Keidar-Levin Y, Schein G (2007) Action bias among
  elite soccer goalkeepers: The case of penalty kicks.
\newblock Journal of Economic Psychology 28: 606--621.
\bibAnnoteFile{bar2007action}

\bibitem{bar2009penalty}
Bar-Eli M, Azar OH (2009) Penalty kicks in soccer: an empirical analysis of
  shooting strategies and goalkeepers' preferences.
\newblock Soccer \& Society 10: 183--191.
\bibAnnoteFile{bar2009penalty}

\bibitem{di2007performance}
Di~Salvo V, Baron R, Tschan H, Calderon~Montero F, Bachl N, et~al. (2007)
  Performance characteristics according to playing position in elite soccer.
\newblock International journal of sports medicine 28: 222.
\bibAnnoteFile{di2007performance}

\bibitem{Correia2011984}
Correia V, Araujo D, Craig C, Passos P (2011) Prospective information for pass
  decisional behavior in rugby union.
\newblock Human Movement Science 30: 984 - 997.
\bibAnnoteFile{Correia2011984}

\bibitem{glickman1998state}
Glickman ME, Stern HS (1998) A state-space model for national football league
  scores.
\newblock Journal of the American Statistical Association 93: 25--35.
\bibAnnoteFile{glickman1998state}

\bibitem{constantinou2010evaluating}
Constantinou A, Fenton N (2010) Evaluating the predictive accuracy of
  association football forecasting systems.
\newblock Technical report.
\bibAnnoteFile{constantinou2010evaluating}

\bibitem{cohea11}
Cohea C, Payton M (2011) Relationships between player actions and game outcomes
  in american football.
\newblock Sportscience 15: 19--24.
\bibAnnoteFile{cohea11}

\bibitem{risen06}
Risen J, Gilovich T, Thaler R (2006) Sudden death aversion: Imagining the worse
  and avoiding it.
\newblock The University of Chicago Booth School of Business (Technical Report)
  .
\bibAnnoteFile{risen06}

\bibitem{RePEc:ucp:jpolec:v:114:y:2006:i:2:p:340-365}
Romer D (2006) {Do Firms Maximize? Evidence from Professional Football}.
\newblock Journal of Political Economy 114: 340-365.
\bibAnnoteFile{RePEc:ucp:jpolec:v:114:y:2006:i:2:p:340-365}

\bibitem{10.2307/2291254}
Albright SC (1993) A statistical analysis of hitting streaks in baseball.
\newblock Journal of the American Statistical Association 88: 1175-1183.
\bibAnnoteFile{10.2307/2291254}

\bibitem{10.1257/jep.5.1.193}
Kahneman D, Knetsch JL, Thaler RH (1991) Anomalies: The endowment effect, loss
  aversion, and status quo bias.
\newblock Journal of Economic Perspectives 5: 193-206.
\bibAnnoteFile{10.1257/jep.5.1.193}

\bibitem{thaler2015misbehaving}
Thaler RH (2015) Misbehaving: The Making of Behavioral Economics.
\newblock WW Norton \& Company.
\bibAnnoteFile{thaler2015misbehaving}

\bibitem{doi:10.1080/00031305.1967.10479847}
Porter RC (1967) Extra-point strategy in football.
\newblock The American Statistician 21: 14-15.
\bibAnnoteFile{doi:10.1080/00031305.1967.10479847}

\bibitem{sackrowitz2000refining}
Sackrowitz H (2000) Refining the point (s)-after-touchdown decision.
\newblock Chance 13: 29--34.
\bibAnnoteFile{sackrowitz2000refining}

\bibitem{nflgame}
 (2012).
\newblock Nfl game center api.
\newblock \url{https://github.com/BurntSushi/nflgame}.
\newblock Accessed: 2016-01-12.
\bibAnnoteFile{nflgame}

\bibitem{barabasi1999173}
Barab\'{a}si A, Albert R, Jeong H (1999) Mean-field theory for scale-free
  random networks.
\newblock Physica A: Statistical Mechanics and its Applications 272: 173 - 187.
\bibAnnoteFile{barabasi1999173}

\bibitem{newman2000mean}
Newman ME, Moore C, Watts DJ (2000) Mean-field solution of the small-world
  network model.
\newblock Physical Review Letters 84: 3201.
\bibAnnoteFile{newman2000mean}

\bibitem{opac-b1127929}
Agresti A (2007) An introduction to categorical data analysis.
\newblock Wiley series in probability and statistics. Hoboken (N.J.):
  Wiley-Interscience.
\bibAnnoteFile{opac-b1127929}

\bibitem{efron93}
Efron B, Tibishirani R (1993) An Introduction to the Bootstrap.
\newblock Chapman and Hall/CRC.
\bibAnnoteFile{efron93}

\bibitem{kunsch89}
K{\" u}nsch H (1989) The jackknife and the bootstrap for general stationary
  observations.
\newblock Annals of Statistics 17: 1217--1241.
\bibAnnoteFile{kunsch89}

\bibitem{sportsnetrank}
Pelechrinis K, Papalexakis E, Faloutsos C (2016) Sportsnetrank: Network-based
  sports team ranking.
\newblock In: ACM SIGKDD Workshop on Large Scale Sports Analytics.
\bibAnnoteFile{sportsnetrank}

\bibitem{cortana}
Tower N (2016).
\newblock Cortana predictions.
\newblock
  \url{https://www.firstscribe.com/blog/bing-predicts-looks-average-in-nfl-week-17-wildcard-weekend-preview/}.
\newblock Accessed: 2016-02-12.
\bibAnnoteFile{cortana}

\bibitem{espn-fpi}
ESPN (2016).
\newblock A guide to nfl fpi.
\newblock
  \url{http://www.espn.com/blog/statsinfo/post/_/id/123048/a-guide-to-nfl-fpi}.
\newblock Accessed: 2016-10-30.
\bibAnnoteFile{espn-fpi}

\bibitem{nfl-expert}
Nerd FF (2015).
\newblock Nfl picks accuracy leaderboard.
\newblock \url{http://www.fantasyfootballnerd.com/nfl-picks/accuracy/}.
\newblock Accessed: 2016-01-12.
\bibAnnoteFile{nfl-expert}

\bibitem{gilovich1985hot}
Gilovich T, Vallone R, Tversky A (1985) The hot hand in basketball: On the
  misperception of random sequences.
\newblock Cognitive Psychology 17: 295 - 314.
\bibAnnoteFile{gilovich1985hot}

\bibitem{kahneman1973psychology}
Kahneman D, Tversky A (1973) On the psychology of prediction.
\newblock Psychological review 80: 237.
\bibAnnoteFile{kahneman1973psychology}

\bibitem{austin2011introduction}
Austin PC (2011) An introduction to propensity score methods for reducing the
  effects of confounding in observational studies.
\newblock Multivariate Behavioral Research 46: 399--424.
\bibAnnoteFile{austin2011introduction}

\end{thebibliography}


\end{document}